\documentclass[preprintnumbers,numbers,sort&compress,
               nofootinbib,3p, 
                            showpacs,
               colorlinks,
               linkcolor=blue,
               citecolor=blue]{elsarticle}

\usepackage{lineno,hyperref}
\modulolinenumbers[5]

\journal{Journal of \LaTeX\ Templates}
 \usepackage{graphicx}% Include figure files
\usepackage{dcolumn}% Align table columns on decimal point
\usepackage{bm}% bold math
\usepackage{hyperref}

\usepackage{graphicx}% Include figure files
\usepackage{dcolumn}% Align table columns on decimal point
\usepackage{bm}% bold math
\usepackage{hyperref}

\usepackage{amsmath}
\usepackage[caption=false]{subfig}
\usepackage{gensymb}% to type the \degree symbol

\newcommand{\exclude}[1]{}

\def\ra{\rangle}
\def\la{\langle}

\newcommand{\commentOut}[1]{}

\newcommand{\be}{\begin{equation}}
\newcommand{\ee}{\end{equation}}
\newcommand{\beq}{\begin{eqnarray}}
\newcommand{\eeq}{\end{eqnarray}}

\graphicspath{{./Figures/}} 

\begin{document}
    \begin{frontmatter}
    
       \title{Ball Lightning as  a profound manifestation of  the Dark Matter physics }
       \author{  Ariel  Zhitnitsky}
       %\email{arz@phas.ubc.ca}
       \address{Department of Physics and Astronomy, University of British Columbia, Vancouver, V6T 1Z1, BC, Canada}
     
      \begin{abstract}
     
Ball lighting (BL) has been observed for centuries.    There are large number of books, review articles, and original scientific papers devoted to different aspects of  BL phenomenon. Yet, the basic features of this phenomenon  have never been explained by known physics.   The  main  problem is the source which could  power the dynamics of the BL.       We advocate an idea that  the dark matter  (DM) in form of the     axion quark nuggets (AQN)    made of standard model  quarks and gluons (similar to the       old idea of the Witten's strangelets) could internally generate the required power. 
% The corresponding macroscopically large object  in form of the AQN behaves as    {\it chameleon}: it  does not interact with the surrounding material in dilute environment and  serves as perfect cold DM candidate.   However,   AQN becomes strongly interacting object in sufficiently dense environment. 
The AQN model was invented long ago without any relation to the BL physics.  It was invented with  a single motivation to explain the observed  similarity     $\Omega_{\rm DM}\sim \Omega_{\rm visible}$ between  visible and DM components.  This relation represents a very  generic feature of this framework,  not sensitive  to any parameters of the construction.    However, with the same set of parameters being fixed long ago this model is capable to address the key elements of the BL phenomenology, including the source of the energy powering the BL events. 
 In particular, we argue that the visible size of BL, its typical life time, the frequency of appearance, etc   are  all consistent with suggested proposal  when BL represents a profound manifestation of the DM physics represented by   the AQN objects. We also argue that some of the Unidentified Aerial Phenomena (UAP) might be closely related to BL events, and therefore also represent profound manifestations of the DM physics within AQN framework. We also formulate a number of specific   possible tests which can refute or unambiguously substantiate  this unorthodox proposal on nature of BL and UAP.  
           
        \end{abstract}


\end{frontmatter}
\section{Introduction}\label{sec:introduction}

The title of this work seemingly includes two contradicting terms: the first one is  ``ball lighting", 
which is very bright luminous  object, though of  unknown nature. The second term of the title is  ``dark matter" (DM)  which is, by definition, must decouple  from the  radiation, as it cannot emit light. Indeed, the BL phenomenon  has been known for centuries,  see recent book \cite{Herbert_Boerner} and review papers \cite{SMIRNOV1993151,SHMATOV2019105115,Rakov_Uman_2003} with large number of references therein.  See also recent review article \cite{hgss-12-43-2021}
with several historical comments on BL observations by scientists and trained professionals. 
 The main goal of this work is to argue that the numerous puzzling  observations of the  ball lighting (BL)    could  be explained from unified viewpoint within   a specific dark matter  model, the  so-called axion quark nugget (AQN) dark matter model  \cite{Zhitnitsky:2002qa},  see   also brief review \cite{Zhitnitsky:2021iwg} for a short overview of the AQN framework.   
 
 We overview the basic ideas of the AQN construction in next section. The main outcome of this construction is  that the AQN   behaves as a chameleon: it serves as a proper DM object in dilute environment, but becomes very strongly interacting object when it hits the stars or planets. Therefore,  the  contradiction in the title is only apparent as the DM in form of the AQNs   become strongly interacting objects in dense environment and can indeed produce  profound and very powerful events  such as BL,
 which is the topic of the present work. 
 
Before we proceed with our  explanations of the    BL phenomenology within AQN framework we should, first of all,  highlight the mysterious properties  of the observations  \cite{Herbert_Boerner,SMIRNOV1993151,SHMATOV2019105115,Rakov_Uman_2003}
 which are impossible  to understand if interpreted in terms of the conventional    physics. 
 In fact a complete failure to understand even the very basic features of the BL phenomenology (such as required power, or passing through a solid glass) enforced the  researchers to look for   possible answers to subatomic physics, well outside the conventional mechanisms considered in the past. In particular,  in ref. \cite{Stephan:2024mau} it was   suggested that the magnetic monopole might be powering the BL, while in ref.  \cite{Ralston:2024xlu} this idea was modified  by adding an electrical  charge into the system by making  the  dyon (magnetic monopole with non-vanishing electric charge).  Our  proposal is also deeply rooted into subatomic physics, but in dramatically different way, as we discuss below. 
 
 \subsection{Brief overview of ball lighting phenomenology}\label{items}
 The first term  of the title is ``ball lighting". Therefore, we have to explain and 
       list the basic  features  of BL. We literally follow  (quote) the review article \cite{Rakov_Uman_2003} which states that a successful model should explain the following observed features: 
       
         \noindent
 (i) ball lightning's association with thunderstorms or with cloud-to-ground lightning;\\
(ii) its reported shape, diameter, and duration, and the fact that its size, luminosity, and appearance generally do not change much throughout its lifetime;\\
(iii) its occurrence in both open air and in enclosed spaces such as buildings or aircraft;\\
(iv) the fact that ball lightning motion is inconsistent with the convective behaviour of a hot gas;\\
(v) the fact that it may decay either silently or explosively;\\
(vi) the fact that ball lightning does not often cause damage;\\
(vii) the fact that it appears to pass through small holes, through metal screens, and through glass windows;\\
(viii) the fact that it is occasionally reported to produce acrid odors and/or to leave burn marks, is occasionally described as producing hissing, buzzing, or fluttering sounds, and is sometimes observed to rotate,
roll, or bounce off the ground.

In addition to these well established features of the BL we want to add few very important recent quantitative measurements suggesting that there is a new scale of the problem on the level $240 \mu m$ \cite{BYCHKOV201669}.
Studies were performed with  optical and scanning microscopes and laser beam probing of the glass that experienced action of 20 cm BL.  Furthermore, the spectrum from BL has been observed  with two slitless spectrographs at a distance 0.9 km, and it includes the soil components (Si I, Fe I, Ca I) as well as the air components (N I, O I) \cite{BL-spectrum-1,BL-spectrum}.  Furthermore, it has been recently claimed that BL radiation must  be accompanied by UV or even X ray emission\cite{STEPHAN201632}.  Therefore we add  few more items to the list:
 
   \noindent
(ix) new scale emerges in the BL problem: $240 \mu m$ which is dramatically different from the visible size of the BL \cite{BYCHKOV201669};\\
(x) spectrum from BL includes lines from soil components (Si I, Fe I, Ca I) as well as the air components (N I, O I) 
\cite{BL-spectrum-1,BL-spectrum}.\\
(xi) spectrum from BL must include UV or/and x ray emission \cite{STEPHAN201632}.
 
We also want to list some BL's characteristics which had been collected for decades and which  play very important role in our discussions. We start with energetic characteristics of the BL. The   energies of the BL events dramatically vary  from case to case and  can be estimated as  follows \cite{SMIRNOV1993151}:
\be
\label{BL-energy}
E_{\rm min}=10^{-0.8\pm 0.2} ~{\rm kJ}, ~~~~ E_{\rm max}=10^{3.2\pm 0.2}~ {\rm kJ} , ~~~ \bar{E}=10^{1.3.\pm 0.2} ~ {\rm kJ}. 
 \ee
The energy density $\epsilon$ also varies for different observations  and has been estimated as \cite{SMIRNOV1993151}:
\be
\label{BL-density}
\epsilon_{\rm min}=10^{-0.6\pm0.5}\frac{J}{\rm cm^3}, ~~~~ \epsilon_{\rm max}=10^{3\pm0.5}\frac{J}{\rm cm^3}, ~~~ \bar{\epsilon}=10^{1.2\pm0.5}\frac{J}{\rm cm^3}.
\ee
 Typical diameter  of the   BL has been estimated as  \cite{SMIRNOV1993151}:
 \be
 \label{BL-size} 
 d=28\pm 4~ {\rm cm}, 
 \ee
 while   life time $\tau$ and   velocity $v_{\rm BL}$ have been estimated from variety of observations as  \cite{SMIRNOV1993151}:
 \be
 \label{BL-time}
 \tau= 9^{+6}_{-4} {\rm ~s},  ~~~ {v}_{\rm BL}\in (0.1-10)  {\rm \frac{m}{s}}, ~~~ \bar{v}_{\rm BL} =4 {\rm \frac{m}{s}}.
 \ee
 Another important  parameter we need for our discussions  is the radiated   power (extracted from studies of the visible frequency bands)  which also strongly varies, and  on average can be estimated as \cite{SMIRNOV1993151}:
\be
\label{BL-power}
P=10^{2.0\pm 0.2} ~W.
\ee
The main obstacle preventing  the  development of  a successful  BL phenomenology  is failure to explain the source of required energy.   Any conventional  theory  (including  any known sources of energy) cannot explain one particular well-established  property of ball lightning: its ability to pass through closed glass windows  \cite{STEPHAN2024106300}. The authors of ref. \cite{BYCHKOV201669} arrived to a similar conclusion by analyzing the glass
damaged by passing BL using the scanning electron microscope as listed in item (ix) above. 
The only physical entities which can easily pass through a few mm of solid glass are subatomic particles, which recently motivated the authors of refs \cite{Stephan:2024mau} and \cite{Ralston:2024xlu} to consider the magnetic monopole as a possible source of BL's energy powering its dynamics.

   \subsection{Brief overview of dark matter}
   The second  term  of the title is ``dark matter". 
    Therefore, we have to briefly  explain  the term  ``dark matter".  From cosmological viewpoint there is a fundamental difference between dark matter
  and ordinary matter (aside from the trivial difference
 dark vs.  visible). Indeed, 
 DM played a crucial role in the formation of the present  structure in the universe.  Without dark matter, the universe would have remained too  uniform to form the galaxies.  
 Ordinary matter could not produce fluctuations to create any significant  structures   because it remains tightly coupled to radiation, preventing it from clustering, until  recent epochs.   
  The key parameter which enters all the cosmological observations is the corresponding cross section $\sigma$ 
  (describing coupling of DM with standard model particles) to mass $M_{\rm DM}$ ratio which must be sufficiently small
to play the role of the DM as briefly mentioned above,  see e.g. recent review  \cite{Tulin:2017ara}:
\be
\label{sigma/m}
\frac{\sigma}{M_{\rm DM}}\ll  1\frac{\rm cm^2}{\rm g}.
\ee
The  Weakly Interacting Massive Particles (WIMP) obviously satisfy to the criteria (\ref{sigma/m}) to serve as DM particles due to their very tiny  cross section $\sigma$ for  a typical mass $M_{\rm WIMP}\in( 10^2-10^3) ~\rm GeV$.
However, the WIMP paradigm  which has been the dominant idea  for the last 40 years has  failed as dozen of dedicated instruments could not find any traces of WIMPs though the sensitivity of the instruments had dramatically improved by many orders of magnitude during the last decades. 

In the present work we consider a  fundamentally  different type of the DM which is in form of  
   macroscopically large composite objects of nuclear density, similar to the Witten's quark nuggets  \cite{Witten:1984rs,Farhi:1984qu,DeRujula:1984axn}.   The corresponding objects   are called the axion quark nuggets (AQN) and behave as    {\it chameleons}: they (almost) 
    not interacting entities      in dilute environment, 
  such that the AQNs may serve as proper DM candidates as the corresponding condition (\ref{sigma/m}) is perfectly satisfied for the AQNs during  the structure formation when  the ratio $\sigma /M_{\rm AQN}\leq 10^{-10} {\rm cm^2}{\rm g^{-1}}$, see (\ref{sigma/M}).  However, the same objects interact very strongly with material when they hit the Earth, or other planets and stars.
  
The main distinct feature of the  AQN model (which plays absolutely crucial role for  the present work) 
in comparison with old Witten's construction 
 is that AQNs  can be made of {\it matter} as well as {\it antimatter} during the QCD transition as a result of the charge segregation   process, see   brief overview \cite{Zhitnitsky:2021iwg}.  This charge segregation mechanism separates quarks  from antiquarks  during the QCD transition in early Universe as a result of dynamics of the $\cal CP$  odd axion field, see more detail explanations and references  in next Sect.\ref{AQN}. This separation of baryon charges lead to formation  of the quark nuggets and anti-quark nuggets with a similar rate.  
  
  \exclude{
  As a result of these processes this model explains  the similarity between the dark matter and the visible matter  densities in the Universe, i.e. $\Omega_{\rm DM}\sim \Omega_{\rm visible}$ without any fine tuning as they both proportional to one and the same fundamental dimensional parameter of the theory, the $\Lambda_{\rm QCD}$ as they both made of the same quarks and gluons. In fact, the main motivation for the AQN construction   \cite{Zhitnitsky:2002qa}, see also a brief review article  \cite{Zhitnitsky:2021iwg} was      to   explain  the observed  similarity between the dark matter and the visible matter  densities in the Universe, i.e. $\Omega_{\rm DM}\sim \Omega_{\rm visible}$.  By no means it was designed to explain BL phenomenology, which is the main topic of this work. The parameters of this model were also fixed long ago
  from different observables, and we shall use the same set of parameters in our analysis of the BL physics. 
 }

 The presence of the antimatter  nuggets in the system implies that there will be  annihilation events leading  to very profound  strong effects when antimatter AQNs hit the Earth. Our claim here is that the  basic features of the BL phenomenology as listed by items (i)- (xi) along with basic BL's characteristics   expressed by  (\ref{BL-energy})-(\ref{BL-power})
   may  naturally emerge  as a result of these  powerful and energetic  annihilation events.  
   
   We reiterate the main claim of this work slightly differently: our proposal is that 
     the source of energy which is powering BL events is related to the annihilation events of the antimatter hidden in form of the AQNs with surrounding atoms and molecules\footnote{In fact, the antimatter as a possible source of the energy powering BL has been discussed long ago \cite{nature-1970,nature-1971}. I thank Karl Stephan for pointing out into these  two old papers.}. The fuel which is powering the BL physics is the antimatter nuggets which had been formed during the QCD transition in early Universe. It has been also shown that the dominant fraction of these  antimatter nuggets mostly survive until present epoch, see brief review on formation mechanism and survival pattern in next Sect. \ref{basics}.
     
     This model was invented long ago to resolve fundamental problems in cosmology, not related in any way to BL phenomena (in contrast with  numerous proposals which were specifically designed to explain different aspects of the BL phenomenology). Nevertheless, this AQN framework  may also shed a light to  another long standing problem which is the nature of the BL. We stress that all parameters for this model were
     fixed long ago  in  our previous applications  to  explain   some mysterious and puzzling observations at galactic  and solar scales, to be reviewed in Conclusion in Sect. \ref{sect:paradigm}.  We keep all these parameters of the AQN model identically the same and we do not attempt to modify them to better fit the observations related to BL phenomenology. 
 
 The presentation of this work is organized as follows. Next Sect. \ref{AQN} represents a brief overview of the AQN construction. In our main Sect. \ref{sect:consistency} we argue that the various of observations  as formulated  above  and collected for decades can be  naturally   explained within AQN framework. In Sect.  \ref{sec:event_rate} we estimate
 the BL event rate and find it is consistent with observations. In Sect. \ref{sect:UAP} we argue that some of the Unidentified Aerial Phenomena (UAP) might be closely related to BL events, and therefore also represent profound manifestations of the DM physics. In our concluding Sect. \ref{conclusion} we suggest a number of specific tests which can substantiate  or refute our proposal on close relation between BL, UAP and dark matter physics within AQN framework. 
   
  \section{The AQN   dark matter  model}\label{AQN}
 We overview  the fundamental  ideas of the AQN model in subsection \ref{basics},  while  in subsections 
 \ref{AQN-dense} and \ref{spallation} we list    some  specific features of the AQNs  relevant to the present work.
 
 \subsection{The basics}\label{basics}
 The original motivation for the AQN model  can be explained as follows. 
It is commonly  assumed that the Universe 
began in a symmetric state with zero global baryonic charge 
and later (through some baryon-number-violating process, non-equilibrium dynamics, and $\cal{CP}$-violation effects, realizing the three  famous  Sakharov criteria) 
evolved into a state with a net positive baryon number.

As an 
alternative to this scenario, we advocate a model in which 
``baryogenesis'' is actually a charge-separation (rather than charge-generation) process 
in which the global baryon number of the universe remains 
zero at all times.   This  represents the key element of the AQN construction.

 In other words,  the unobserved antibaryons in visible sector  in this model comprise 
dark matter being in the form of dense nuggets of antiquarks and gluons in the  colour superconducting (CS) phase.  
The result of this ``charge-separation process'' are two populations of AQN carrying positive and 
negative baryon number. The global  $\cal CP$ violating processes associated with the so-called initial misalignment angle $\theta_0$ which was present  during 
the early formation stage,  the number of nuggets and antinuggets 
  will be different.
 This difference is always an order-of-one effect irrespective of the 
parameters of the theory, the axion mass $m_a$ or the initial misalignment angle $\theta_0$.

The presence of the antimatter nuggets in the AQN  framework is an inevitable and the direct consequence of the 
    $\cal{CP}$ violating  axion field  which is present in the system during the  QCD time. As a result of this feature      the DM density, 
    $\Omega_{\rm DM}$, and the visible    density, $\Omega_{\rm visible}$, will automatically assume the  same order of magnitude densities  
    $\Omega_{\rm DM}\sim \Omega_{\rm visible}$  irrespective  to the parameters of the model, such as the axion mass $m_a$. 
 This feature represents a generic property of the construction   \cite{Zhitnitsky:2002qa,Zhitnitsky:2021iwg}.

% As we already mentioned 
% the AQN construction in many respects is 
%similar to the Witten's quark nuggets, see  \cite{Witten:1984rs,Farhi:1984qu,DeRujula:1984axn}. 
This type of DM  is ``cosmologically dark'' as a result of smallness of the parameter  (\ref{sigma/m})  relevant for cosmology.  
This numerically small ratio scales down many observable consequences of an otherwise strongly-interacting DM candidate in form of the AQN nuggets.  
Indeed, for a typical AQN parameters the relevant ratio  assumes the following numerical value:
\be
\label{sigma/M}
\frac{\sigma}{M_N}\sim \frac{\pi R^2}{M_N}\sim    10^{-10}{\rm \frac{cm^2}{g}},  
\ee
where for numerical  estimate we use parameters  from  Table \ref{tab:basics} to be discussed below. It obviously satisfies the cosmological constraint (\ref{sigma/m}).  

Another important characteristic is   the AQN flux  which we need in what follows to estimate the BL occurrency  rate. It can be estimated  as follows \cite{Lawson:2019cvy}:
 \be
\label{Phi1}
\frac{\rm d \Phi}{\rm d A}
=\frac{\Phi}{4\pi R_\oplus^2}  =  4\cdot 10^{-2}\left(\frac{10^{25}}{\langle B\rangle}\right)\rm \frac{events}{yr\cdot  km^2},
\ee
 where $R_\oplus=6371\,$km is the radius of the Earth, while $\langle B\rangle \approx 10^{25}$ is a typical baryon charge of the AQNs 
  from Table \ref{tab:basics} to be discussed below,   and  $\Phi$ is the total hit rate of AQNs on Earth \cite{Lawson:2019cvy}:
\be
\label{Phi}
\Phi
\approx \frac{2\cdot 10^7}{\rm yr}  
 \left(\frac{\rho_{\rm DM}}{0.3{\rm\,GeV\,cm^{-3}}}\right)
\left(\frac{v_{\rm AQN}}{220~ \rm km ~s^{-1}}\right)
\left(\frac{10^{25}}{\langle B\rangle}\right), 
\ee
where $\rho_{\rm DM}$ is the local density of DM within Standard Halo Model (SHM).   

\exclude{
However, there are several additional elements in the AQN model in comparison with the older well-known and well-studied  {theoretical} constructions \cite{Witten:1984rs,Farhi:1984qu,DeRujula:1984axn}. First of all, the nuggets can be made of {\it matter} as well as {\it antimatter} during the QCD transition
as we already mentioned.   Secondly, there is an additional stabilization factor for the nuggets provided by the axion domain walls which  are copiously produced  during the   QCD  transition. This additional element    helps to alleviate a number of  problems with the original Witten's  model. In particular, a first-order phase transition is not a required feature for nugget formation as the axion domain wall (with internal QCD substructure)  plays the role of the squeezer. 
}
 We refer to the original papers   \cite{Liang:2016tqc,Ge:2017ttc,Ge:2017idw,Ge:2019voa}    devoted to the specific questions  related to the nugget's formation, generation of the baryon asymmetry, and  survival   pattern of the nuggets during the evolution in  early Universe with its unfriendly environment.  In particular, the absolute stability of the AQNs 
in vacuum is a result of the construction when the energy per baryon charge in   the quark-matter nuggets is smaller than in the baryons (from hadronic phase) making   up the visible portion of the Universe.

 % The only comment we would like to make here is that in this work   we take the  agnostic viewpoint, and assume that such nuggets made of  antimatter are present in our Universe today   irrespective to their   formation mechanism. 
 This construction  is consistent with all presently available cosmological, astrophysical and terrestrial  constraints as long as  the average baryon charge of the nuggets is sufficiently large as we review  below.
Precisely the presence of the antimatter nuggets make the AQNs to be very strongly interacting objects as a result of annihilation of the AQNs with surrounding material when AQNs  hit  stars or planets.  
%A direct consequence of this feature, given that the total baryon charge of the Universe is zero, is that DM density, $\Omega_{\rm DM}$, and the baryonic matter density, $\Omega_{\rm visible}$, will automatically assume the  same order of magnitude  $\Omega_{\rm DM}\sim \Omega_{\rm visible}$ without any fine tuning.

 In the AQN scenario  the DM density, $\Omega_{\rm DM}$ representing the matter and anti-matter nuggets, and the visible    density, $\Omega_{\rm visible}$, will automatically assume the  same order of magnitude densities  $\Omega_{\rm DM}\sim \Omega_{\rm visible}$   as they both proportional to one and the same fundamental dimensional parameter of the theory, the $\Lambda_{\rm QCD}$. Therefore, the AQN  model,   by construction, actually resolves two fundamental problems in cosmology  (explains the baryon asymmetry of the Universe, and the presence of   DM with proper density  $\Omega_{\rm DM}\sim \Omega_{\rm visible}$) without necessity  to fit any parameters of the model.

 \begin{table*}
\captionsetup{justification=raggedright}
	%\centering % used for centering table
	\begin{tabular}{cccrcc} % centered columns (4 columns)
		\hline\hline
		  Property  && \begin{tabular} {@{}c@{}}{ Typical value or feature}~~~~~\end{tabular} \\\hline
		  AQN's mass~  $[M_N]$ &&         $M_N\approx 16\,g\,(B/10^{25})$     \cite{Zhitnitsky:2021iwg}     \\
		   baryon charge constraints~   $ [B]  $   &&        $ B \geq 3\cdot 10^{24}  $     \cite{Zhitnitsky:2021iwg}    \\
		   annihilation cross section~  $[\sigma]$ &&     $\sigma\approx\kappa\pi R^2\simeq 1.5\cdot 10^{-9} {\rm cm^2} \cdot  \kappa (R/2.2\cdot 10^{-5}\rm cm)^2$  ~~~~     \\
		  density of AQNs~ $[n_{\rm AQN}]$         &&          $n_{\rm AQN} \sim 0.3\cdot 10^{-25} {\rm cm^{-3}} (10^{25}/B) $   \cite{Zhitnitsky:2021iwg} \\
		  survival pattern during BBN &&       $\Delta B/B\ll 1$  \cite{Zhitnitsky:2006vt,Flambaum:2018ohm,SinghSidhu:2020cxw,Santillan:2020lbj} \\
		  survival pattern during CMB &&           $\Delta B/B\ll 1$ \cite{Zhitnitsky:2006vt,Lawson:2018qkc,SinghSidhu:2020cxw} \\
		  survival pattern during post-recombination &&   $\Delta B/B\ll 1$ \cite{Ge:2019voa} \\\hline
	\end{tabular}
	\caption{Basic  properties of the AQNs adopted from \cite{Budker:2020mqk}. The parameter $\kappa$ in Table  is introduced to account  for possible deviation from geometric value $\pi R^2$ as a result of ionization of the AQNs due to interaction with environment. The ratio $\Delta B/B\ll 1$ in the Table implies that only a small portion $\Delta B$  of the total (anti)baryon charge $  B$  hidden in form of the AQNs get annihilated during big-bang nucleosynthesis (BBN), Cosmic Microwave Background (CMB), or post-recombination epochs (including the galaxy and star formation), while the dominant portion of the baryon charge survives until the present time.  } % title of Table
	\label{tab:basics}
\end{table*}

 The strongest direct detection limit\footnote{Non-detection of etching tracks in ancient mica gives another indirect constraint on the flux of   DM nuggets with mass $M> 55$g   \cite{Jacobs:2014yca}. This constraint is based on assumption that all nuggets have the same mass, which is not the case  for the AQN model.} is  set by the IceCube Observatory's,  see Appendix A in \cite{Lawson:2019cvy}:
\be
\label{direct}
\la B \ra > 3\cdot 10^{24} ~~~[{\rm direct ~ (non)detection ~constraint]}.
\ee
The basic idea of the constraint (\ref{direct}) is that IceCube  with its surface   area $\rm \sim km^2$  has not detected any events during  its 10 years of observations. In the estimate  (\ref{direct})
it was assumed that the efficiency of  detection of a macroscopically large nugget is 100$\%$ which excludes AQNs with small baryon charges 
$\la B \ra < 3\cdot 10^{24}$ with   $\sim 3.5 \sigma$ confidence level.

The authors of Ref. \cite{SinghSidhu:2020cxw} considered a generic constraint for the nuggets made of antimatter (ignoring all essential  specifics of the AQN model such as quark matter  colour superconducting (CS) phase of the nugget's core, see \cite{Alford:2007xm} for  introduction of the  CS phases.). Our constraints (\ref{direct}) are consistent with their findings including the Cosmic Microwave Background (CMB) and Big Bang Nucleosynthesis (BBN), and others, except the constraints derived from    the so-called ``Human Detectors". 
As explained in \cite{Ge:2020xvf}
  the corresponding estimates of Ref. \cite{SinghSidhu:2020cxw} are   oversimplified   and do not have the same status as those derived from CMB or BBN constraints.  

     \begin{figure}[h]
	\centering
	\captionsetup{justification=raggedright}
	\includegraphics[width=0.8\linewidth]{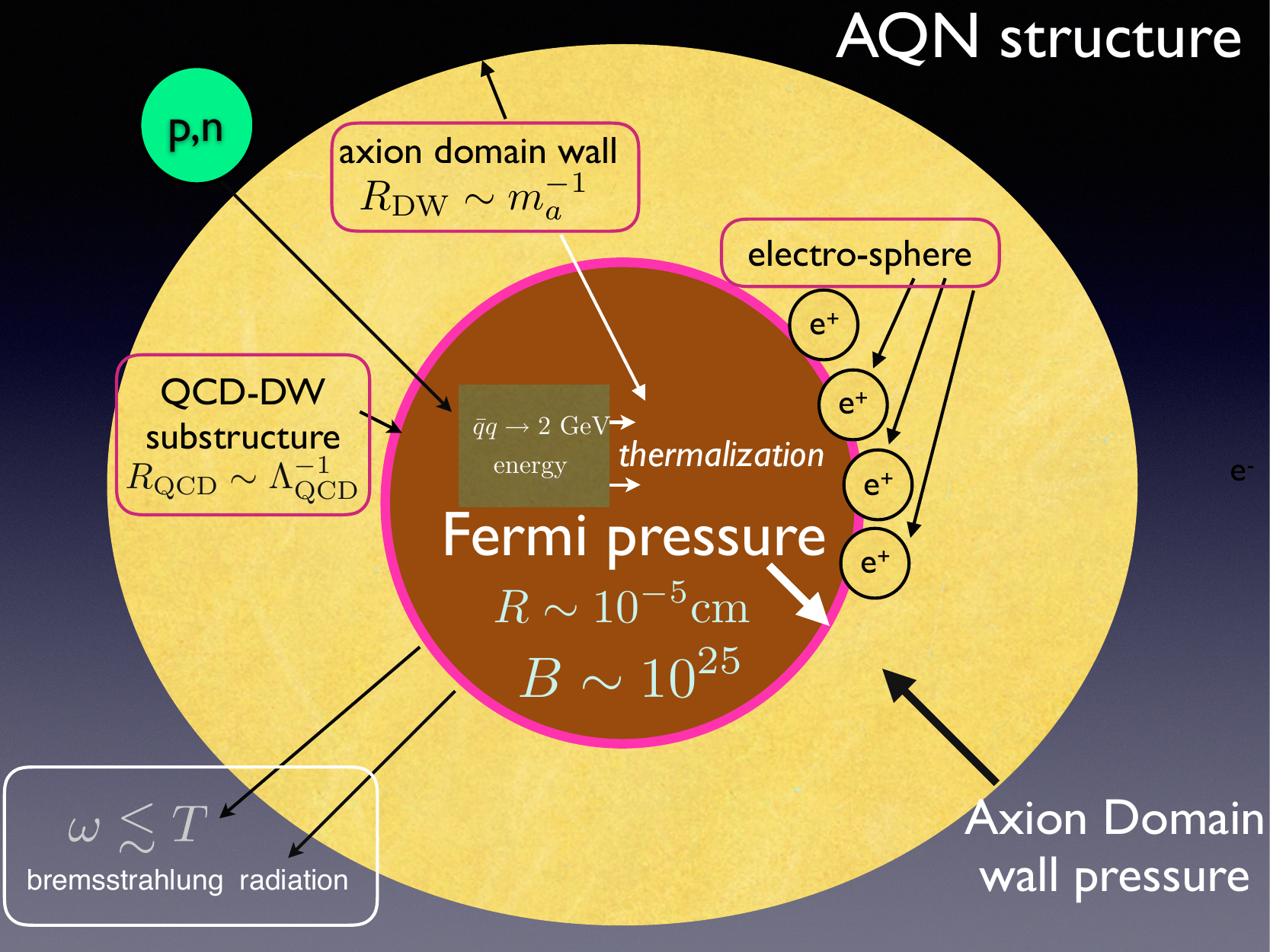}
		\caption{AQN-structure (not in scale), adopted from \cite{Zhitnitsky:2022swb}. The dominant portion of the energy $\sim 2$ GeV produced as a result of  a single  annihilation process inside the anti-nugget is released in form of the bremsstrahlung radiation with frequencies $\omega\leq T$, see description and notations in the main text.}
\label{AQN-structure}
\end{figure}

  We draw the AQN-structure on Fig \ref{AQN-structure}, where we use typical parameters from the  Table\,\ref{tab:basics}. There are several   distinct length scales of the problem: $R\sim 10^{-5}$ cm represents the size of the nugget filled by dense quark matter with total baryon charge $B\sim 10^{25}$ in CS phase. Much larger scale  $R_{\rm DW}\sim m_a^{-1}$  describes the axion DW   surrounding the quark matter. The axion DW has the QCD substructure surrounding the quark matter and  which has typical width of order $R_{\rm QCD}\sim 10^{-13} \rm cm$. Finally, there is always electro-sphere which represents a  very generic feature of quark nuggets, including the Witten's original construction. In case of antimatter-nuggets the electro-sphere comprises the positrons.  The  typical size of the electrosphere is order of $10^{-8} \rm cm$.

We conclude this brief review subsection with Table\,\ref{tab:basics} which summarizes the basic features and parameters of the AQNs.
     Important point here is that
 only a small portion $\Delta B\ll B$  of the total (anti)baryon charge $  B$  hidden in form of the AQNs get annihilated during long evolution of the Universe, while  
 the dominant portion of the baryon charge survives until the present time.
 % Independent analysis  \cite{SinghSidhu:2020cxw}   and  \cite{Santillan:2020lbj}  also support our original claims as cited in the Table\,\ref{tab:basics} that the anti-quark nuggets survive the Big Bang Nucleosynthesis (BBN),  Cosmic Microwave Background (CMB) and recombination epochs. 
 
 %The large mass of  the nuggets along with their small sizes also  implies that  the direct head on AQN-AQN collisions are extremely rare events and  do not modify our estimates for $\Delta B/B\ll 1$.  

 \subsection{When the AQN hits the Earth}\label{AQN-dense}
       The    computations of the AQN-visible matter interaction originally have been carried out in \cite{Forbes:2008uf}
 in application to the galactic   environment    with a typical density of surrounding   baryons of order $n_{\rm galaxy}\sim   {\rm cm^{-3}}$ in the galaxy. We review  these computations  below with few additional elements which must be implemented in case of propagation of the AQN in denser  environment such as Earth's  atmosphere with $n_{\rm atm}\sim  10^{21} {\rm cm^{-3}}$.

When the AQN enters the region of the baryon density $n $ the annihilation processes start and the internal temperature increases. 
  A typical internal temperature  $T$ of  the  AQN   can be estimated from the condition that
 the radiative output   must balance the flux of energy onto the 
nugget \cite{Forbes:2008uf}:
\be
\label{eq:rad_balance}
    F_{\rm{tot}} (T) (4\pi R^2)
\approx \kappa\cdot  (\pi R^2) \cdot (2~ {\rm GeV})\cdot n \cdot v_{\rm AQN},  
\ee 
where $n$ represents the baryon number density of the surrounding material, and $F_{\rm{tot}}(T) $ is total  surface emissivity, see below. 
The left hand side accounts for the total energy radiation from the  AQN's surface per unit time   while  
 the right hand side  accounts for the rate of annihilation events when each successful annihilation event of a single baryon charge produces $\sim 2m_pc^2\approx 2~{\rm GeV}$ energy. 
 %If the environment is represented by neutral atoms and molecules the interaction of the AQNs with environment can be approximated by the geometrical cross section $ \pi R^2$ for macroscopically large object of size $R$. 
% However, if the surrounding material is highly ionized the effective cross section $  \pi R_{\rm eff}^2$ could be dramatically larger than the geometric value $ \pi R^2$ due to the long range Coulomb interaction as the AQN assumes a large negative charge at sufficiently high temperature $T$. 
 The factor $\kappa$  in (\ref{eq:rad_balance}) accounts    for large theoretical uncertainties related to the annihilation processes 
of the (antimatter)  AQN  colliding with surrounding material.

        The total  surface emissivity due to the bremsstrahlung radiation from electrosphere at temperature   $T$ has been computed in \cite{Forbes:2008uf} and it is given by 
\begin{equation}
  \label{eq:P_t}
  F_{\rm{tot}} \approx 
  \frac{16}{3}
  \frac{T^4\alpha^{5/2}}{\pi}\sqrt[4]{\frac{T}{m}}\,,
\end{equation}
 where $\alpha\approx1/137$ is the fine structure constant, $m=511{\rm\,keV}$ is the mass of electron, and $T$ is the internal temperature of the AQN.  
 One should emphasize that the emission from the electrosphere is not thermal, and the spectrum is dramatically different from blackbody radiation.
    
    As we mentioned above,  the thermal properties in (\ref{eq:rad_balance}), (\ref{eq:P_t}) were originally applied to the study of the emission from AQNs from the Galactic Centre, where a nugget's internal temperature is very low, $T\sim$ eV. 
When the nuggets propagate in the Earth's atmosphere, the AQN's internal temperature starts to rise up to $\sim 20$ keV or so.    From (\ref{eq:rad_balance})  one can estimate a typical internal nugget's temperature in the Earth atmosphere as follows:
 \be
 \label{T}
 T\sim 20 ~{\rm keV} \cdot \left(\frac{n_{\rm air}}{10^{21} ~{\rm cm^{-3}}}\right)^{\frac{4}{17}}\left(\frac{ \kappa}{0.1}\right)^{\frac{4}{17}},
 \ee 
 where  typical density of surrounding baryons is   $n_{\rm air}\simeq 30\cdot N_m\simeq 10^{21} ~{\rm cm^{-3}}$, where 
 $N_m\simeq 2.7\cdot 10^{19}  ~{\rm cm^{-3}}$ is the molecular density  in atmosphere when each molecule contains approximately 30 baryons. 
Thus, in the atmosphere  the internal nugget's temperature $T\approx$ 20 KeV. 
Similar   temperature $T\approx 20$ keV had been  previously used in \citep{Zhitnitsky:2020shd}   to explain the unusual Cosmic Ray like events  observed by Telescope Array Collaboration (so-called ``TA bursts"), and in  \cite{Zhitnitsky:2022swb} to explain ``Exotic Events" recorded by  the  AUGER   collaboration. 
%The   $T\sim 200$ keV had been previously used in \cite{Zhitnitsky:2021qhj} to explain unusual clustering events observed by the HORIZON -10T collaboration and in \cite{Liang:2021rnv} to explain two anomalous events    with non-inverted polarity  recorded by  \textsc{ANITA}. 
%In what follows, we use $T\approx 400 $ keV as the benchmark temperature for the nuggets exiting the Earth's surface in the upward direction. 
%A precise  computations  of $T$ from the first principles is a very difficult problem. Therefore, we opted to keep $T$ as the phenomenological parameter, which will be determined by observational constraints. 

Another  feature   which is relevant for our  present studies is the ionization  properties of the AQN. Ionization, as usual,   occurs in a system   as a result of  the high internal temperature $T$. In our case of the AQN characterized by temperature (\ref{T})     a large number of weakly bound positrons $\sim Q$ from the electrosphere   get excited and can easily leave the system. The corresponding parameter $Q$ can be estimated as follows: 
\be
  \label{Q}
Q\approx 4\pi R^2 \int^{\infty}_{0}  n(z, T) d z\sim \frac{4\pi R^2}{\sqrt{2\pi\alpha}}  \left(m T\right)   \left(\frac{T}{m}\right)^{\frac{1}{4}} , ~~
  \ee
 where $n(z, T)$ is the local density of positrons at distance $z$ from the nugget's surface, which has been computed in the mean field approximation  in \cite{Forbes:2008uf} and has the following form
\begin{equation}
\label{eq:nz0}
n(z, T)=\frac{T}{2\pi\alpha}\frac{1}{(z+\bar{z})^2}, ~~ ~~\bar{z}^{-1}\approx \sqrt{2\pi\alpha}\cdot m  \cdot \left(\frac{T}{m}\right)^{\frac{1}{4}}, ~~~~  \end{equation}
where $\bar{z}$ is the integration constant is chosen to match the Boltzmann regime at sufficiently  large $z\gg \bar{z}$. Numerical 
studies ~ \cite{Forbes:2009wg}  support the approximate analytical  expression (\ref{eq:nz0}).

  Numerically, 
the number of weakly bound positrons can be estimated from (\ref{Q}) as follows:
\be
\label{Q1}
Q\approx 3\cdot 10^{11}   \left(\frac{T}{20~ \rm keV}\right)^{\frac{5}{4}}  \left(\frac{R}{2.25 \cdot10^{-5} \rm cm}\right)^{2}.
\ee
 These positrons from electrosphere being in the equilibrium  (when the AQN experiences a  relatively small  annihilation rate in dilute galactic environment) will normally occupy very thin layer    around the  AQN's quark core as computed in \cite{Forbes:2008uf,Forbes:2009wg}. However, in our case when the AQN enters the Earth's atmosphere    a large number of non-equilibrium processes 
 (such as generation of the shock wave  resulting from large Mach number) are expected to occur. 
 In what follows  we   assume that, to first order, that the  finite portion of positrons  $\sim Q$   leave the system as a result of the complicated processes mentioned above, in which case    the AQN as a system   acquires a negative  electric  charge $\sim -|e|Q$  and  get   partially  ionized as a macroscopically large object of mass $M\simeq m_pB$. 
The ratio $eQ/M\sim 10^{-14} e/m_p$ characterizing  this object  is very tiny. 
However, the charge $Q$ itself is sufficiently large  being capable to   attract the positively charged  ions from air (for example during the thunderstorm).

  \subsection{AQN spallation}\label{spallation}
  One more feature of the AQN propagating in Earth's atmosphere   (which  plays an important role for the present work) is as follows. As mentioned in Sect.\ref{basics} the AQN is absolutely stable object because the energy per unit baryon charge in CS phase (AQN's core) is smaller than for nucleons. However, some external strong impact and large energy injection (due to sudden annihilation events within the AQN's quark core)  may disintegrate AQN when a  small chunk from  the original AQN material (in form of the anti-quark-matter) is separated from the original parent AQN.  In a sense, it is very similar to well known and well studied    spallation effect which is very common phenomenon in nuclear physics. We coin the corresponding secondary particles as AQN$_s$, where subscript $s$ stands for  {\it  s}econdary particle or 
  {\it s}pallation. 
  
  The secondary AQN$_s$ are obviously not absolutely stable objects as the key element for the stability, the axion domain wall with its QCD substructure, see Fig.\ref{AQN-structure}, cannot remain in the system after spallation. This should be contrasted with conventional spallation effect in nuclear physics   when both, the original nucleus and the secondary particles are absolutely stable objects (with respect to strong interactions) as they both propagate in the same hadronic (low chemical potential, low temperature) phase\footnote{The secondary AQN$_s$ can be thought as the metastable states, similar to supercooled or superheated liquid droplets when large chunk of matter suddenly appears in a ``wrong"  environment (temperature and pressure) not supporting its phase. }, while AQN's core is assumed to be in CS phase characterized by sufficiently large  chemical potential (large pressure), see phase diagram in ref. \cite{Ge:2019voa}.

    The secondary AQN$_s$ could be much smaller in size than   AQN. The corresponding fragment   really represents  very small chunk of the original (anti-matter) material. As we discuss in next Sect.\ref{sect:consistency} the 
    secondary AQN$_s$ will be identified with BL events as conjecture (\ref{eq:proposal}) states. Precisely these chunks of antimatter material will play the role of engines powering the BL dynamics in this proposal. 
         These objects  should have typical sizes $B_{AQN_s}\approx 10^{15}$ (to be discussed below),  which is 10 orders of magnitude smaller than   original AQN with typical baryon charge $B\approx 10^{25}$.     %  cannot be computed from the first principles 
   % as even the phase diagram (as well as the corresponding excitations, the quasi-particles)  is not well understood for CS phase which assumed to be realized in AQN. 
 
   In our discussions which follow we assume that the  spallation is likely to occur as the most important ingredient required for spallation, a high fraction of ionized particles in surrounding area, is present during the thunderstorms. As we discuss below, the highly ionized environment dramatically increases the effective strength of interaction of the AQN with surrounding material expressed by parameter $\kappa$ in (\ref{T}), which drastically  increases the internal temperature leading to spallation, see Sect. \ref{sect:charge} with more comments on this matter. The spallation dynamics, 
    the size distribution  of the secondary AQN$_s$ and other related questions have not been worked out yet (detail discussions on  formation mechanism
    for original AQN can be found in ref. \cite{Ge:2019voa}). The corresponding  questions are well beyond  the scope  of the present work. We take agnostic view in this work on  spallation dynamics. We just assume that the spallation   occurs and we use a typical size (or what is the same, the baryon charge)  of  the secondary AQN$_s$   to fit the BL observations. This is the only input parameter to be used in the present work, while all other observables  relevant for BL physics will be derived from this single input parameter.
%   In  next Sect.\ref{sect:consistency} we fix a typical size of the secondary AQN$_s$ from numerous observations  of the BL if identification (\ref{eq:proposal}) is assumed.  However, 
%One should emphasize that the variation of the secondary AQN$_s$  in size after spallation could be very large, which is in fact consistent with dramatic variations in energetic characteristics as numerous observations (\ref{BL-energy}) suggest. 

To conclude this overview section on AQN framework  one should mention here that this  model with the same set of parameters to be used in the present work may explain a number of   puzzling and mysterious observations 
%(when un-explained  excess of emission  has been observed in many frequency bands) 
which cannot be understood  as   conventional astrophysical phenomena.  
%The corresponding events observed on  Earth have been already mentioned at the end of Sect. \ref{sec:introduction}.  
In particular, there are many   mysterious observations which  occur at dramatically  different scales at  very  different  cosmological eras, which also might be related to AQN-induced phenomena.  It includes    BBN epoch, dark ages, as well as  galactic and  Solar    environments. It also includes a number of puzzling and mysterious CR-like events which could be also related to AQN-induced phenomena,   see  concluding sections 
\ref{sect:tests} and \ref{sect:paradigm} for the  details and references.

  \section{Proposal (\ref{eq:proposal}) confronts  the observations}\label{sect:consistency}
   Our proposal  can be formulated as follows. The secondary particles (after   spallation) in form of the  antimatter AQN$_s$ are  identified with Ball Lightning events, i.e.
    \be
    \label{eq:proposal}
  \rm   secondary ~AQN_s ~events ~~~\equiv ~~~ Ball~ Lightning ~events. 
     \ee
The main goal of this section is to argue that the various of observations  as formulated in Introduction \ref{sec:introduction} collected for decades can be  naturally   explained  (though very often on a  qualitative level)    if one accepts the proposal
 (\ref{eq:proposal}).    
 
 \subsection{Source of the energy powering BL}\label{sect: energy}
 The source of the energy in the AQN framework is obviously the antimatter annihilation with surrounding material.
 Every annihilation event of a single baryon produces approximately 2 GeV energy.    We fix the 
   basic parameter of the proposal  (\ref{eq:proposal}) by fixing the total amount of antimatter in form of the secondary $AQN_s$. According to (\ref{BL-energy}) the typical energy for BL events vary from $1$ kJ to $10^3$ kJ.  
   The mean energy for BL is estimated in  \cite{SMIRNOV1993151} as $2\cdot 10^2 \rm kJ$. To simplify our estimates below we use $10^2 kJ$ as a typical  average energy for BL events. Therefore, we fix the amount of antimatter hidden in form of the $AQN_s$ accordingly:
   \be
   \label{s-energy}
 \rm   10^2 kJ\approx 10^{15} GeV ~~~ \rightarrow ~~~ B_{\rm AQN_s}\approx 10^{15},     ~~~~  R_{\rm AQN_s}\approx\left(\frac{B_{\rm AQN_s}}{B}\right)^{\frac{1}{3}}R\approx 10^{-8} \rm cm
\ee
   where we use conversion factor $J= 0.6\cdot 10^{10} \rm GeV$.
 One should emphasize that the chunks of the $AQN_s$ could have very different sizes. Therefore,  a wide  window 
 for the BL energy distribution (as extracted  from observations,    see Fig 5 in  \cite{SMIRNOV1993151})     could be easily accommodated in  our proposal (\ref{eq:proposal}) to fit the data.  
 However, we stick with the average value (\ref{s-energy}) in what follows for simplicity our qualitative analysis. One should also add that a large variation in    energies of the BL events (\ref{BL-energy}) strongly suggest that the source of the energy is unlikely to be related to any fundamental subatomic particles when one should expect a similarities for  different events. In contrast, the proposal  (\ref{eq:proposal}) is based on complex classical macroscopical system with variety of sizes which is perfectly consistent with broad distribution of the energy scales  (\ref{BL-energy}). 
 
 \subsection{BL is electrically charged}\label{sect:charge}
 Now we want to argue that $AQN_s$  will carry the electric charge $Q_{\rm AQN_s}$ after spallation.
 The corresponding numerical value can be easily estimated by assuming that the internal temperature after spallation remain the same as for original AQN (\ref{Q1}). Therefore, one can use the rescaling to arrive to the estimate:
   \be
   \label{s-charge}
  \frac{Q_{\rm AQN_s}}{Q}\approx\left(\frac{B_{\rm AQN_s}}{B}\right)^{\frac{2}{3}}\approx  \left(\frac{10^{15}}{10^{25}}\right)^{\frac{2}{3}}\approx 2\cdot 10^{-7} ~~~ \rightarrow ~~~ Q_{\rm AQN_s}\approx 6\cdot 10^4, 
     \ee
 where we   used the feature that the $Q$ charge is the surface effect, while the baryon $B$ charge is the volume effect. 
 The presence of the charge in the system will play a key role in the dynamics to be discussed below. First of all, presence of the charge $Q$ in the system in the original AQN implies that the AQN can attract the positively charged ions from the air such that the rate of annihilation may dramatically increase.  This effect can be described by drastic increase of the phenomenological coefficient $\kappa$ in (\ref{eq:rad_balance}) such that effective temperature $T$   also suddenly increases when AQN enters the region of a highly ionized gas. As a consequence, the spallation phenomenon may be much more efficient (and likely to occur) in this case.
 
  The presence of large density of ionized particles in air is known to occur during the thunderstorms in thunderclouds (even without strikes). 
   Therefore, the association of the BL with thunderstorms, see item (i) from Sect.\ref{items}, has its direct explanation within our proposal  due to a sudden and  dramatic increase of $\kappa$ (and corresponding internal temperature $T$) as mentioned above. These dramatic changes  may lead to spallation effect which consequently results in formation of the secondary particles $ {\rm AQN_s}$ which are identified with BL according to (\ref{eq:proposal}). 
  
  Furthermore, the secondary particles $ {\rm AQN_s}$   are also charged according to  (\ref{s-charge}).  One should emphasize that this estimate for $Q_{\rm AQN_s}$  refers to  the internal (bound) charge which are localized very close to the quark core with size (\ref{s-energy}). The induced charge due to other processes, see below could be much greater.
  %  In fact,  the  $AQN_s$   charge (\ref{s-charge}) plays the role of a trigger when  a dramatic 
 %  increase of the effective charge of the entire $AQN_s-$ system 
  %Furthermore, as we argue below, a large number density of the ionized atoms from air will be localized in vicinity of  the core of    
  The presence of the electric  charge in the system  is also consistent with observations which suggest that BL gets attracted (due to the induced charges) to the metallic  surfaces, wires, antennas  \cite{SMIRNOV1993151,Rakov_Uman_2003}.  
  
  \subsection{Spectral properties of the BL radiation. Size of   BL(in visible frequency bands).}\label{sect:radiation}
  
  The spectral properties  of the AQN emission has been studied previously as reviewed in Appendix \ref{sect:spectrum}. 
  The key feature of this radiation is very broad spectrum (in contrast with black body radiation) up to cutoff frequency  which occurs around  $\omega\approx  T$. What happens to these keV photons emitted by  the  $  AQN_s$ (identified with BL) with internal core temperature $T\approx 6$ keV  according to (\ref{T-spallation})?
  
  The dominant mechanism of absorption for such energetic X rays is the atomic photoelectric effect, see Fig 33.15 for 
  $Z=6$ (Carbon) in PDG \cite{PDG}. For the Oxygen (Z=8) and Nitrogen (Z=7) the effect   is  slightly higher  due to very strong dependence on $Z$ \cite{berestetskii1982quantum}: 
  \be
  \label{photoeffect}
  \sigma_{\rm photo-effect}\approx \frac{32\sqrt{2}\pi}{3}r_0^2Z^5\alpha^4\left(\frac{m}{\omega}\right)^{\frac{7}{2}}, ~~~~~~~r_0\equiv \frac{\alpha^2}{m^2}=2.8\cdot 10^{-13} \rm cm.
  \ee
   Based on this cross section the mean free path $\lambda $ for 6 keV photons can be estimated as
    \be
   \label{lambda1}
   \lambda^{O,N} \rm (6 ~keV)\approx 10 ~  cm,   ~~~~~~~~~~~~~~ [ to~ be~ identified~ with~visible~ size ~of~ BL]. 
   \ee
     where we used experimental data for $Z=6$ (Carbon),  see Fig 33.19 and rescaled for the Oxygen (Z=8) and Nitrogen (Z=7). 
   
   The atomic photoelectric effect is accompanied by electron emission. The life time of a free electron in air  is  very short, around $0.1\mu \rm s$, see e.g. \cite{Gurevich_2001} such that free electron will be quickly absorbed by atoms in air on very short distances around $10^{-4} \rm m$, much shorter than (\ref{lambda1}). These excited and ionized atoms and molecules (made of  Oxygen and Nitrogen along with many other elements from soil) will emit visible light which is observed as radiation coming from BL according to our proposal (\ref{eq:proposal}). Therefore, we identify the   size of BL (in visible frequency bands) with mean free path $\lambda$ as stated in (\ref{lambda1}). This scale is perfectly consistent with observations (\ref{BL-size}).
   
   Few comments are in order:\\
   1. The mean free path $\lambda$ introduced above is highly sensitive to the frequency   of radiation (or what is the same, the internal temperature). Therefore, even a minor variation in internal temperature of the $AQN_s$ can dramatically modify the mean free path. To illustrate this feature we estimate  $\lambda$ for 10 keV photons: 
   \be
   \label{lambda}
   \lambda^{O,N} \rm  (10 ~keV)\approx 70 ~  cm, ~~~~~~~~~~~~~~ [ to~ be~ identified~ with~ visible~size ~of~ BL]
   \ee

   Our main point with this illustration is that the visible size of the BL in this proposal (\ref{eq:proposal}), being identified with $\lambda$,  is not related to a basic energetic characteristic (\ref{BL-energy}) which could dramatically  vary (three orders of magnitude or more) from one event to another.  Rather, the variation in size of the BL is related to very small variation in internal temperature $T$;\\
   2.   This picture of emission of the visible light is consistent with observation that average density of BL is the same as average density of the surrounding air because the quark matter core with $M_{\rm AQN_s}\approx 10^{-9} \rm g$ is negligible in comparison with weight of air in volume $\lambda^3$. The observed visible light from BL in this proposal is obviously not related to hot plasma, nor to convective behaviour of a hot gas, see  (item iv) from Sect.\ref{items}.  This proposal also naturally explains the observation that BL normally moves horizontally as the average density of the BL object is the same as the air density at the same   temperature\footnote{The description of the spectrum in terms of very high temperature $T\sim 10^4$K as presented in  \cite{BL-spectrum-1,BL-spectrum}, see next item 3, is a matter of convenience to describe a highly ionized system. In fact there is no real temperature in the system as there is no  thermal equilibrium  when the temperature could be defined in the system. The thermal equilibrium  is obviously cannot be achieved in BL system. The same comment obviously applies to the lightning phenomenon as it is a highly non- equilibrium processes. The black body radiation which normally characterizes the system being in the thermal equilibrium is not present in data  \cite{BL-spectrum-1,BL-spectrum}. The eyewitness  also suggest that 
   there is no any heat associated with    BL \cite{SMIRNOV1993151,SHMATOV2019105115,Rakov_Uman_2003}.}; \\
   3.  This picture of emission of the visible light is consistent with observation  that the spectrum contains soil components (Si I, Fe I, Ca I) (item x) from Sect.\ref{items}.    This is because the soil components have much larger $Z$ (and therefore, the cross section (\ref{photoeffect}) is much greater). As a result, even a tiny amount of these components in air (e.g. in form of the dust particles) could generate very   strong intensity lines associated with these elements, which is consistent with observations \cite{BL-spectrum-1,BL-spectrum};\\
   4.  From the estimate (\ref{lambda1}) of BL size in the visible frequency bands one can estimate the average energy density    for our specific parameters (\ref{s-energy})  for  BL propagating in air as follows:
   \be
   \label{s-density}
  \rm  \epsilon^{O,N}\equiv \frac{(B_{\rm AQN_s} \cdot m_p)}{ (\lambda^{O,N} \rm  )^2 \bar{L}}\approx \frac{10^2 kJ}{(10~ cm)^2\cdot (40~ m)}\approx  0.25  \frac{J}{cm^3},  ~~~~~ \bar{L}\approx \bar{v}_{\rm BL}\cdot \tau\approx 40 ~\rm m,
   \ee
   where $\bar{L}$ is an average length for  BL's path with parameters from (\ref{BL-time}). 
   This estimate  is consistent with   value (\ref{BL-density})   extracted from observations in visible frequency bands;\\
    5. This picture of emission  is consistent with observation  that BL emits UV or x ray radiation along with visible light as discussed above. In fact, UV and x rays are originated from the the core of the $AQN_s$, in contrast with visible light 
   which is a secondary process in the AQN framework as described above.  This  picture is perfectly consistent with   (item xi) from Sect.\ref{items} when the presence of UV or x rays from BL had been directly observed \cite{STEPHAN201632}.
   The presence of the UV or x rays from BL is also supported our estimate for the total power of emission which suggests that the power in visible light represents only a small fraction of the total power, see next Sect. \ref{sect:life};\\
   6. Annihilation of the baryons is always accompanied by annihilation of the electrons from atoms with positrons from AQN's electrosphere. The corresponding total energy injection  per single event (MeV scale) due to $e^+e^-$ annihilation  is negligible ($\sim 10^{-3}$) in comparison with GeV scale due to hadron's annihilations. However the emission of the 0.511 MeV photons from $e^+e^-$ annihilation may play a crucial role in understanding of items viii and  xi from Sect.\ref{items}. This is because the mean free path for such  energetic photons in air  is very large, $\rm \lambda (0.5 ~MeV)\sim 10~ m$ such that these  photons can ionize the surrounding space and could be responsible for ionizing radiation.   A number of  observed phenomena which could be related to    ionizing radiation  from BL were  reported  in \cite{SHMATOV2019105115}. 
      
        \subsection{BL life time. The power of radiation. The BL's    internal  size.}
      \label{sect:life}
      The internal size of quark matter of the $AQN_s$ is  determined by  the quark matter core $R\approx 10^{-8} \rm cm$ from (\ref{s-energy}). However, the effective size of the quark matter material with surrounding positrons is dramatically larger than quark matter core itself (similar to atoms with typical size $a\approx 10^{-8} \rm cm$  being much larger than 
      the size of a nuclei $\sim 10^{-13} \rm cm$). 
      
      We define  the effective radius  $R_{\rm eff}$ as the scale where positrons from electrosphere remain to be strongly bound to the quark's core  at the internal temperature $T\approx 6$ keV.  The corresponding scale can be estimated from the   condition that the binding energy of the positrons is approximately equal to internal temperature $T\approx 6~ \rm keV$. This gives 
    $  R_{\rm eff}\approx 10^{-6} \rm cm$, see (\ref{R_eff}) in Appendix \ref{sect:spectrum} for numerical estimates.  The corresponding scale should be treated as internal structure of the system, in  contrast with the scales (\ref{lambda1}) and (\ref{lambda2})  which should be treated as environment-dependent scales. 
    
    The significance  of the scale   $R_{\rm eff}$ is related to the fact that this scale determines the rate of the antimatter annihilation    (and therefore BL's life time in the AQN framework) hidden in form of the $AQN_s$. To be more precise, the rate of annihilation of the baryon charge due to head on collisions of the air molecules with anti-matter  $AQN_s$ can be estimated as follows:
  \be
\label{eq:rate}
     \frac{d B}{dt}\approx   -(\pi R_{\rm eff}^2) \cdot  n_{\rm air} \cdot v_{\rm air},  
\ee 
where we assume that the successful  annihilation events  represent a finite fraction of order one for all collisions. Another  finite fraction of collisions  are elastic scattering events when molecules  of air scatter without annihilation.
 In formula  (\ref{eq:rate}) we take a typical density of surrounding baryons as   $n_{\rm air}\simeq 30\cdot N_m\simeq 10^{21} ~{\rm cm^{-3}}$,with  
 $N_m\simeq 2.7\cdot 10^{19}  ~{\rm cm^{-3}}$ being  the molecular density  in atmosphere when each molecule contains approximately 30 baryons. The rate of annihilation as given by expression (\ref{eq:rate}) in all respects is      similar in spirit to the right hand side of eq. (\ref{eq:rad_balance}) in our  estimation for  the energy injection rate for the AQN when it just enters the earth's atmosphere.  
 
 The  $v_{\rm air}$ (typical velocity of molecules in air) can be estimated from condition 
 \be
 \label{eq:v_air}
   \frac{m_{\rm air}v_{\rm air}^2}{2}\approx \frac{3}{2}T_{\rm air} ~~ \rightarrow ~~ v_{\rm air}\approx 5\cdot 10^4 \rm \frac{cm}{s},
 \ee
 where $T_{\rm air}\approx 300 K$ is air  temperature.
   Collecting all the numerical   factors together we arrive to   estimate:
       \be
\label{eq:rate1}
     \frac{1}{B_0}\frac{d B(t)}{dt}\approx    -  \frac{1}{B_0} {\rm    \frac{1.5\cdot 10^{14}~ baryons}{s}}\left(\frac{R_{\rm eff}}{10^{-6}\rm cm}\right)^2\approx -\frac{1}{\tau}, ~~~  \tau\approx 6.7 s
\ee 
      where we inserted $B_0^{-1}$ for normalization with $B_0=B(t=0)$ being  defined as initial (anti)baryon charge of the $AQN_s$ with   its effective radius as  estimated in  (\ref{R_eff}). In reality the time scale  (\ref{eq:rate1}) for our benchmark value for  $B_0$ as given by (\ref{s-energy}) is slightly longer 
      as we ignored in our numerical estimates the elastically  scattered  events (which obviously decrease the rate (\ref{eq:rate}) and consequently increase the estimate for $\tau$) as mentioned above. 
      Few comments are in order:\\
      1. The time scale which appears in  (\ref{eq:rate1})  is perfectly consistent with observations according to (\ref{BL-time}). This is a highly nontrivial consistency check for entire proposal (\ref{eq:proposal}) 
      because two observed parameters (total energy of the BL (\ref{s-energy})  and its life time (\ref{eq:rate1}) ) are unambiguously connected in this proposal. It is very hard to imagine any other mechanism when these two very different entities  are tightly connected and agreed with observed values;\\
      2. The power of radiation (total power) can be estimated as 
      \be
      \label{eq:power}
      P_{\rm tot}\approx \frac{B_{\rm AQN_s}\cdot m_p}{\tau}\approx \rm \frac{10^2 kJ}{\tau}\approx 10 ~kW,
       \ee
       which is consistent with (\ref{BL-power}) extracted from studies in  the visible frequency bands. Our estimate 
       (\ref{eq:power}) suggests that the emission in visible light could be only a small fraction of the total power generated by BL;\\
        3. Formula (\ref{eq:rate1}) holds as long as  portion  $f(t)\equiv B(t)/B_0$ represents a finite fraction of the initial baryon charge of order one. This is because    for  very small  $f(t)\ll 1$ some dramatic changes in rate of annihilation may occur which consequently may result in explosion  instead of smooth and slow decay  determined by the  time scale $\tau$, see next comment. This behaviour   is consistent with item (v) from the list in  Sect.\ref{items};\\
      4. The  explosion occurs if the the annihilation rate (\ref{eq:rate1}) suddenly increases due to some external impacts or as a result of successful simultaneous annihilation of  a large number of baryons from air when internal temperature must instantaneously increase to equilibrate heating and cooling processes.  In this case the  time scale (\ref{eq:rate1}) suddenly and dramatically  decreases resulting in very intense flash of broad band radiation   and consequent formation of an acoustic shock wave.  This would appear as an explosion of BL when all remaining antimatter in the $AQN_s$ get annihilated at once. The properties of the resulted shock wave are different from conventional chemical or nuclear explosions but similar   to the ones studied in \cite{Budker:2020mqk};\\
      5. The size  of the BL in visible bands  during a smooth evolution in this framework is determined by the photon's mean-free path as explained in Sect. \ref{sect:radiation}. This scale    is not very sensitive to a slow decreasing of the (anti)baryon charge in the quark core during the BL evolution, as it is determined by (almost) constant internal temperature $T$ according to formula (\ref{T-spallation}) from Appendix \ref{sect:spectrum}. This conclusion  is in agreement  with observations from item (ii) from the list in  Sect.\ref{items}.

      \subsection{BL passing through glass windows. The BL's   new scale of the problem.}
      \label{sect:glass}
       The authors of ref. \cite{BYCHKOV201669}   with a help of optical and  scanning microscopes and laser beam probing the glass,  have found the traces which  are  left by 20 cm BL passing through the window glass.  The authors  discovered a cavity of 0.24 mm  diameter, see Fig.3 in that paper.  
      The authors  correctly interpreted this event as an undeniable   evidence of  a ``material" nature of BL. The scale of this 
      cavity is dramatically different from the  20 cm scale   of the BL  as observed at visible frequency bands. 
      
      In the AQN framework the emergence of this new scale (0.24 mm in comparison with 20 cm scale as  observed in visible light) has a very natural explanation. Indeed, the mean free path for a  similar  energy photons in Si (12 keV in Si instead of 6 keV in air, see relevant comment in footnote\footnote{\label{crossing}Precise estimation of the internal temperature when 
      the  $AQN_s$ crosses   the interface between very different environments   is very difficult  technical problem.  In particular,  the internal temperature obviously should  increase  in comparison with temperature in air (\ref{T-spallation}) when the interface is crossed.  However, the computation of this increase is   hard  to carry out. In particular, the  thermal equilibration in the $AQN_s$ electrosphere for this short passage of the interface is unlikely to hold, which obviously complicates the problem.   We account for this and related effects by increasing the effective temperature from 6 keV to 12 keV which appears in (\ref{lambda2}) to fit the observed value. A proper procedure to account for this and related effects is to solve the problem for  the $AQN_s$ dynamics when it crosses the interface between  air and  glass which are characterized by  dramatically different densities and atomic compositions. The corresponding computation is well beyond  the scope of the present work.})  is dramatically shorter in comparison with O and N atoms from air. It can be extracted from the same  Fig 33.19 from PDG \cite{PDG} which gives the following estimate:
       \be
   \label{lambda2}
 \rm   \lambda^{Si} \rm  (12 ~keV)\approx 0.25 ~  mm, ~~~~~~~~~~~~~~ [ to~ be~ identified~ with~ size ~of~a~ cavity~ in ~ glass]
   \ee
      where we use $\rm \rho (Si)=2.3 g/cm^3$ for the estimate. 
      % Estimate (\ref{lambda2})  is consistent with item ix from the list in Sect.\ref{items}.  
      The  estimate (\ref{lambda2}) is very instructive as it shows a dramatic difference 
      between two scales (\ref{lambda1})  and (\ref{lambda2}).  A proper  computation of the internal temperature  (and corresponding photon's energy) entering the estimate (\ref{lambda2}) is very hard technical problem, see footnote \ref{crossing}. However, our generic   claim that the scale  (\ref{lambda2}) must be dramatically smaller than the scale (\ref{lambda1}) is a  solid qualitative prediction of the  framework as it is entirely determined by  the differences in photon's mean free paths for two very different environments with similar x-rays energies.  
      %The similarity of the observed scale  (0.24 mm) and the estimate (\ref{lambda2}) is very encouraging
      The estimate  (\ref{lambda2}) shows the consistency of the entire framework when the scale of the BL phenomena is determined by the interaction of the $AQN_s$ core with environment, rather than by its internal structure. As we discussed in previous Sect.  \ref{sect:life} the internal size of the  $AQN_s$ core   is dramatically smaller than scale (\ref{lambda2}). 
      
      The energy density injected at the instant when  BL passing through the window glass of width $  l\approx \rm 2mm$ can be estimated from (\ref{lambda2}) as follows:
       \be
   \label{s-density-glass}
    \epsilon^{Si}\equiv \frac{(\Delta B_{\rm AQN_s} \cdot m_p)}{ (\lambda^{Si}    )^2 l}\approx \frac{10^2 J}{(\rm 0.25 ~mm)^2 (2~mm)}\approx    10^3 \frac{\rm kJ}{\rm cm^3},  ~~~~~~~ \Delta B_{\rm AQN_s}= (\pi R_{\rm eff}^2 l) \cdot  n_{\rm glass} \approx 10^{-3} B_{\rm AQN_s},
   \ee
   where $\Delta B_{\rm AQN_s}$ is amount of antimatter being  annihilated during  BL's  passage  through the window glass of width $  l\approx \rm 2mm$. In this estimate 
    we use the same   $  R_{\rm eff}\approx 10^{-6} \rm cm$ which enters (\ref{eq:rate}) from previous Sect. \ref{sect:life}. 
   For numerical estimates we use the baryon number density of glass $n_{\rm glass}= \rho_{\rm glass}/m_p\approx 1.4\cdot 10^{24} \rm cm^{-3}$. 
  
   Important  point here is that $  \rm  \epsilon^{Si}$ 
     is dramatically greater than a similar estimate for BL propagating in air as given by (\ref{s-density}).
This enormous energy density is obviously more than sufficient to melt the glass in small volume of size $ (\lambda^{Si}    )^2 l$. The process of the glass melting in the AQN framework can be thought as (almost) instantaneous event when  the 12 keV photons    radiate along $AQN_s$  path with 
 area of size $(\lambda^{Si}    )^2 $ and length $l$. This (almost) instant process   ends after $AQN_s$ passes through glass windows on time scale $  l/v_{BL} \approx 0.5 \rm ms$, after which BL returns to its  previous original  size 20 cm observed for BL   propagating  in air. Such short changes in size in time scale of order $0.5 \rm ms$ of course cannot be noticed by a human eye (the threshold is about 20 ms).
In fact all witnesses report that no any   changes occur during  the course of BL passing through glass windows. 
   Few comments are in order:\\
   1. In the AQN  framework the BL passing through glass windows (or any other surfaces) is a very natural effect.
   Indeed, the   available energy density of the BL  crossing a solid material could be  very large according to (\ref{s-density-glass}).
      When the BL crosses the window the emitted (from $AQN_s$ core) photons will be  localized    on a scale of order (\ref{lambda2}). As a result of this ``focusing" effect the energy density   assumes enormous  value (\ref{s-density-glass}) in form of a very short pulse with time scale of order 0.5ms.  This enormous energy density  is sufficient to melt essentially any material;\\
   2. This picture of passing BL through the window is consistent with studies of ref. \cite{BYCHKOV201669}
   where    ``one can assume that the heating of the glass was carried out by a powerful pulse of electromagnetic radiation" (it is a direct quote);\\
   3. The emergent scale of the problem (\ref{lambda2})  is not related in anyway to the internal structure of the BL itself
    which  was discussed in previous  Sect. \ref{sect:life}. 
   Rather, the scale (\ref{lambda2})  emerges due to   interaction of the $AQN_s$ with environment (window glass) where the mean free path $ \rm   \lambda^{Si}$ is relatively short. It should be contrasted with our   discussions of propagation of BL in air in Sect. \ref{sect:radiation} where  dramatically different scale (\ref{lambda1}) emerges.\\

           \subsection{How does BL emerge (in form of the $AQN_s$)  after spallation? }\label{sect:stopping}
      As we mentioned in Sect. \ref{spallation} the parent 
      AQN may disintegrate (as a result of some external impacts, see below) 
     when one or several smaller pieces  
         consisting the   original AQN material (in form of the anti-quark-matter) get separated.
         The phenomenon in many respects is  similar to well known process in nuclear physics. We coined the corresponding secondary particles as AQN$_s$ and identify them with BL  (\ref{eq:proposal}). As we mentioned previously, the  spallation effect  may become highly  efficient  during the thunderstorms in thunderclouds (even without strikes)
 due to presence of large density of ionized particles and dramatic increase   of the annihilation rate which consequently results in sudden increase of the internal temperature $T$.           
   Therefore, the association of the BL with thunderstorms characterized by high density of ionized particles,  has its  natural (though very qualitative) explanation within our proposal.  
            
        The AQN itself represents the DM macroscopical object  which enters the Earth's atmosphere  with         typical velocity $v_{AQN}\sim 10^{-3}c$. It may cross the Earth by  loosing only tiny portion of the   momentum and   antimatter material along its path due to a very large baryon number $B\approx 10^{25}$ carried by the system, as reviewed in Sect.\ref{basics}. The situation with the secondary $AQN_s$ is dramatically different  because of its much smaller baryon charge $B_s\approx 10^{15}$ in which case a  complete annihilation of the antimatter material in atmosphere (or earth's surface) becomes  inevitable. 
        
   The question we address in this subsection is as follows. How long does it take for $AQN_s$ to slow down from typical velocity $10^{-3}c$ to essentially zero velocity at the earth's surface when BL is normally observed? To estimate the corresponding length scale $L$ (when stopping occurs) we observe  that the elastic head-on collision of $AQN_s$ with  a single baryon charge leads to  decrease of  the initial $AQN_s$ momentum by amount $\sim 2m_p 10^{-3}c$. In case of annihilation or non-head on collision the decrease of momentum is numerically smaller. However, for a simple estimation we can assume that the amount of material from air   in a cylinder of  radius $R_{\rm eff.}$ and length $L$ must be the same order of magnitude  as the baryon charge $B_s\approx 10^{15}$  for the $AQN_s$ to   loose its huge initial velocity, i.e.
     \be
\label{eq:stop}
  B_s\sim    (\pi R_{\rm eff}^2) \cdot  n_{\rm air} \cdot L~~~~~\rightarrow ~~~~  L\sim 3~\rm km,
\ee 
where   $R_{\rm eff}\approx 10^{-6} \rm cm$ is  the internal effective size of the $AQN_s$ which has been  used  previously in Sect. \ref{sect:life}.
 Few comments are in order:\\
 1. We consider the numerical value for $L$  to be  a very reasonable   estimate. Indeed, the scale (\ref{eq:stop}) corresponds to a typical size  of the thunderclouds. Therefore, the $AQN_s$ can reach the earth's surface after spallation in thunderclouds by loosing its huge original  velocity to become BL with very low velocity at the surface where it is normally    observed. A typical stoppage time $\tau_s$, when $AQN_s$ looses its 99\% of its momentum,  can be  estimated as $\tau_s     \approx \rm 2~s$ which is  slightly   shorter than BL's life time $\tau$ from (\ref{eq:rate1}). At this velocity the BL becomes  observable in visible frequency bands. A quantitative estimate for this stoppage time $\tau_s     \approx \rm 2~s$ is given in Appendix \ref{sect:stoppage};\\
 2. One can explicitly see that  very small    $AQN_s$ with  $B_s\ll 10^{15}$ cannot survive a several kilometres  journey from thunderclouds to the earth's surface as they get completely annihilated long before they reach the surface.  This could be a simple   explanation (within AQN framework) for the  well established  feature that BL has a lower energetic bound   (\ref{BL-energy}).  In the AQN framework this bound emerges as a result of identification (\ref{eq:proposal})  when    two entities (energy $E_{\rm BL}$ and baryon charge $B_s$) are tightly linked  in our proposal: $E_{\rm BL}\approx B_s (2  \rm GeV)  $;\\
 3. The estimate (\ref{eq:stop}) also shows why extremely large $B_s\gg 10^{15}$ have never been observed as BL events. Indeed, the observed maximum   for  $E_{\rm max}$ is,  at most,  two orders of magnitude above the average  value according to (\ref{BL-energy}).
 In the AQN framework this feature is explained as follows. Very large values of  $B_s\gg 10^{15}$ imply  that the stoppage distance $L$ must be much longer  in comparison with  our estimate (\ref{eq:stop}). Therefore, the secondary $AQN_s$ with very large $B_s\gg 10^{15}$,   if they are formed, will hit the earth's surface with very high velocities   and get completely annihilated only in deep underground regions, in contrast with BL which assume very low velocity 
 near the surface. As a result, such energetic events with $B_s\gg 10^{15}$ are less likely to be observed in comparison with typical BL\footnote{In fact, there are many recorded     events, which are classified as ``Pseudo-meteorites  events" when a meteor-like event is observed, but no actual physical meteorite  is found in the area, see  \ref{sect: ELMA} for references and details.};\\
  4. One should emphasize that the spallation is not a mechanism of production of anti-quark material powering BL. 
 Antiquark nuggets had been produced during the QCD transition in early Universe and survived until present epoch, as reviewed in Sect. \ref{basics}.
 Spallation is a secondary  phenomenon when small portion 
  of this anti-quark material disintegrated from the original anti-matter AQN. 
 In other words, spallation is not a production of engine powering the BL. Rather, this engine in form of the antimatter 
 (we observed today in form of BL) had been produced during the QCD transition  in early Universe.

        \subsection{Summary. Consistency of the proposal  (\ref{eq:proposal})  with observed features  from Sect.\ref{items}}
      \label{sect:summary}
      In subsections  \ref{sect: energy}-\ref{sect:stopping} above we argued that  our proposal  (\ref{eq:proposal})  is perfectly consistent with observed items  from Sect.\ref{items}. However, all our explanations and estimates 
      were scattered in   the text.
      The goal here, for consistency and uniformity of presentation,    is to summarize and collect all of the items in the same order, one  by one  with a precise reference to a  specific and detail estimate given in the text. The observed features  from Sect.\ref{items} include:\\
      (i) BL's association with thunderstorm is discussed in Sects.  \ref{sect:charge} and \ref{sect:stopping} with specific estimate (\ref{eq:stop}) of a distance BL propagates from thunderclouds where it was formed to the surface where it is normally observed. The basic reason for 
      thunderclouds to play a key role in BL formation is the generation of the AQN's internal negative electric charge which 
      dramatically  increases the interaction with positively charged ions from surrounding area during  thunderstorms; \\
      (ii) Typical size of BL in the visible frequency bands is discussed in Sect. \ref{sect:radiation} with specific estimates as given by (\ref{lambda}) and (\ref{lambda1}). Lifetime of BL is discussed in Sect. \ref{sect:life} with specific estimate for $\tau$ as given by (\ref{eq:rate1}). The arguments suggesting that there should be no strong time-variation  of these parameters  throughout  the BL's time evolution is presented  in item 4 in Sect. \ref{sect:life};\\
      (iii) Most of BL events are likely to occur  in open air as BLs (in the AQN framework) propagate to earth's surface from thunderclouds. However, BL can easily cross a glass or any other material and continue to propagate in enclosed spaces such as buildings  or aircraft, see 
   Sect.  \ref{sect:glass} with corresponding discussions and estimation for the cavity size for glass (\ref{lambda2}). 
   Similar estimates  are applicable  for any other materials, including metals in  case of an aircraft;\\
   (iv) Average density of the BL is the same as surrounding air, see item 2 in Sect.  \ref{sect:radiation}.
   This is because the observed radiation from BL is not associated in any way with heating of the air inside the visible part of BL. The emission in visible frequency bands from BL has dramatically different nature as explained in details in Sect.  \ref{sect:radiation} with specific estimation (\ref{lambda1}) of the visible portion of BL;\\
   (v) A typical lifetime of BL is discussed in Sect. \ref{sect:life} with specific estimate for $\tau$ as given by (\ref{eq:rate1}).
   This estimate assumes a smooth evolution. In some cases (resulting from some external impacts) the explosion may occur as mentioned in item 3  in Sect. \ref{sect:life};\\
   (vi) In case of a still  evolution the lifetime is determined by formula (\ref{eq:rate1}) from Sect. \ref{sect:life}.
   This formula holds for smooth propagation of BL  in air when its entire original energy is  released in steady  way in form of the x rays and visible light without much damage to  surrounding area;\\
   (vii) The process of BL  crossing  through metal screens or  glass is described in  Sect.\ref{sect:glass} with specific estimate (\ref{lambda2}) for size of a cavity in glass as a result of such passage. A typical time for such passage is estimated on the level of $ 0.5~ \rm ms$ such that this fast variation in BL's size cannot be noticed by a human eye;\\
   (viii) The radiation from the $AQN_s$ is very broad band in nature. In particular, it includes MeV photons along with x rays, see item 6 in Sect. \ref{sect:radiation}. Furthermore, the $AQN_s$ carries internal negative charge which could be much larger in value than original initial charge (\ref{s-charge}) after spallation. The atomic photo-effects described in Sect.  \ref{sect:radiation} may also ionize surrounding air. 
  All these phenomena   may produce a number of effects described in (viii), including   acrid odors and others phenomena due to ionizing radiation as reported  in \cite{SHMATOV2019105115};\\
  (ix)  In the AQN framework the emergence of this new scale (0.24 mm) as reported in ref. \cite{BYCHKOV201669} can be naturally explained, see Sect. \ref{sect:glass}.  The main point is that the scale (\ref{lambda2})  emerges due to   interaction of the $AQN_s$ with environment (window glass). This new scale is not related to the internal structure of the BL;\\
  (x)  the spectrum contains soil components (Si I, Fe I, Ca I) according to  \cite{BL-spectrum-1,BL-spectrum}.     This is because the soil components have much larger $Z$ such that the  cross section (\ref{photoeffect}) is much greater than for dominant air components O and N. As a result, even a tiny amount of these soil components in air   could produce    strong intensity lines associated with these elements, see  item 3 in Sect.\ref{sect:radiation};\\
  (xi) The direct observations explicitly show \cite{STEPHAN201632} that the spectrum from BL must include UV or/and x ray emission. This observation  is perfectly consistent with the picture of emission advocated in this work, see items  5 and 6 from Sect. Sect.\ref{sect:radiation}. 
  
  \subsection{Concluding comments on proposal   (\ref{eq:proposal})}\label{sect:comments}
  
  We conclude this section with the following very generic comments.
  The AQN model was suggested long ago to resolve some fundamental problems in cosmology, see Sect.\ref{AQN}. 
  The corresponding AQN parameters (including spectral properties of radiation, etc) were also worked out  long ago 
  to address many puzzles and mysteries mostly related to observed excesses of radiation at different frequencies bands at different cosmological and astrophysical scales, see Sect.\ref{sect:paradigm} for a brief  review.  The AQN model  was not designed to  address  the BL physics.  In our estimates in this work we use exactly the same set of parameters extracted from our cosmological studies   to test   bold unorthodox proposal   (\ref{eq:proposal}) relating the BL and DM physics.
  
   We produced a number of estimates relating   the  observed BL parameters   listed in Sect. \ref{items}  with AQN parameters.     All our  estimates   in this section are based 
  on a single input parameter (\ref{s-energy}), the baryon charge (or what is the same, the energy of BL)
    of the $AQN_s$ identified with BL
  according to the proposal (\ref{eq:proposal}). In particular, the visible size of BL (\ref{lambda1}), its typical life time (\ref{eq:rate1}), the cavity size (\ref{lambda2})   are determined by 
 incorporating this single input normalization parameter (\ref{s-energy})  with variety of well known physical observables, such as photo-effect cross section for different atoms,  the  density of the environment, mean free paths,   etc. 
 
 It is a  highly nontrivial consistency check that all our estimates  of the BL  characteristics   in this section assume very reasonable values being consistent with observations relating BL and DM physics in AQN framework. Indeed, the rate of annihilation determines the internal temperature of the $AQN_s$ according to (\ref{T-spallation}), which consequently determines the spectrum, which (through the atomic photo-effect) determines the mean free path and  the size of the BL in visible frequency bands (\ref{lambda1}) in air. The same parameters  determine the  life time of the BL according to  (\ref{eq:rate1}). It strongly supports the basic idea formulated as (\ref{eq:proposal})
 that the anti-matter nuggets, representing the DM objects    in empty space, in fact become the engines powering the   BL   physics and producing   profound effects when these objects enter the earth's atmosphere.
 
 This consistency check in fact extends to the  next Sect \ref{sec:event_rate} where we  demonstrate  that the frequency of appearance of BL is consistent with flux of DM objects in form  of  the AQNs.  In this case  the basic normalization factor is the dark matter density $\rm \rho_{DM}=0.3~ GeV\cdot cm^{-3}$ extracted  from numerous cosmological studies. It turns out that precisely this factor $\rm \rho_{DM}$ determines the frequency of appearance of BL, which   further strengthen   the  proposal  (\ref{eq:proposal}) relating BL and DM physics.  
                  
\section{Frequency of appearance} \label{sec:event_rate} 

We start with overview of the recent  analysis  \cite{Stephan:2024mau,STEPHAN2022105953} for frequency of BL appearance.
The corresponding lower bound flux has been estimated as \cite{Stephan:2024mau}:
\be
\label{Phi_BL}
\frac{\rm d \Phi_{BL}}{\rm d A d \Omega} > 1.75\cdot 10^{-24}\rm \frac{events}{ cm^2 \cdot sr\cdot s}.
\ee
The author of ref.  \cite{Stephan:2024mau} argues that the actual frequency of BL  is in fact much greater than 
lower bound (\ref{Phi_BL}). However, it is hard to estimate. We would like to represent the lower bound (\ref{Phi_BL}) using more appropriate (for rare events)  units 
\be
\label{Phi_BL1}
\frac{\rm d \Phi_{BL}}{\rm d A} >  4\cdot 10^{-6}\rm \frac{events}{ km^2 \cdot   yr},
\ee
where we insert factor $ 2\pi$ into  (\ref{Phi_BL1})  as the solid angle  to account for all  directions from entire sky. 
One can formulate the following question. It is known that BL are associated with lightning, see item (i) in Sect. \ref{items}. Therefore, one can argue that the lightning should play an important role in formation of the BL\footnote{In fact, there are many models suggesting that BL is formed as a result of lightnings.}. Than,  why the rate (\ref{Phi_BL1}) is much  lower than an average frequency of conventional lightnings in the continental U. S., which is   about $24~ \rm km^{-2}~yr^{-1}$? We rephrase the same question in a different way: what is so special about very rare lightning events which produce BL (\ref{Phi_BL1}) in comparison with vast majority of lightning  events which do not lead to BL?

Now we turn to another side of our story, the Dark Matter.  In the AQN framework the corresponding AQN flux is proportional to the dark matter number density $  n_{\rm DM} \propto \rho_{\rm DM}/\langle B\rangle$.  It is convenient to represent the DM flux as given by (\ref{Phi1}) and (\ref{Phi})   as follows: 
  \be
  \label{Phi2}
\frac{\rm d \Phi}{\rm d A}
=\frac{\Phi}{4\pi R_\oplus^2}  =  4\cdot 10^{-2}\left(\frac{\rho_{\rm DM}}{0.3{\rm\,GeV\,cm^{-3}}}\right)
\left(\frac{v_{\rm AQN}}{220~ \rm km ~s^{-1}}\right)\left(\frac{10^{25}}{\langle B\rangle}\right)\rm \frac{events}{yr\cdot  km^2},
\ee
where we assumed the standard   halo model with the local dark matter density being  $\rho_{\rm DM}\simeq 0.3\,{\rm  {GeV} {cm^{-3}}}$ and canonical galactic wind  $v_{\rm AQN}\simeq 220~ \rm km ~s^{-1}$. The number density of the AQNs is very tiny: $n_{\rm AQN}\simeq \left(\frac{\rho_{\rm DM}}{m_p\langle B\rangle} \right)$ as it is suppressed by factor $B^{-1}$. This should be contrasted with canonical type of WIMPs with typical mass $\sim 10^2 \rm~ GeV$ in comparison  with AQN mass $\sim 10^{25} \rm ~GeV$. Conventional DM detectors designed for WIMP searches  are obviously useless to study the DM  in form of the AQNs as there will be no any events for million of years. The Cosmic-Ray (CR) labs, on other hand, may record the AQN-induced events, and we shall comment on this with relation to recording some unusual CR -like events during the thunderstorms   in Sect.\ref{sect:tests}.
%This is precisely the main reason  for  the nuggets to behave as the  cold dark matter particles. 

Now we are ready to estimate the flux for BL within AQN framework when AQN hits a thunderstorm area, which consequently leads to spallation and formation  of the BL according to the proposal (\ref{eq:proposal}).
Assuming that every AQN which hits the area under thunderclouds (where ionization is high and spallation is likely to occur) produces a single secondary $AQN_s$ we arrive to the following  estimate for the BL flux within AQN framework: 
 \be
  \label{BL-flux}
\frac{\rm d \Phi^{AQN}_{BL}}{\rm d A}\approx \frac{\rm d \Phi}{\rm d A}\cdot {\cal{F}}
 \approx  4\cdot 10^{-2} {\cal{F}} \left(\frac{\rho_{\rm DM}}{0.3{\rm\,GeV\,cm^{-3}}}\right)
 \rm \frac{(BL~events)}{yr\cdot  km^2},
\ee
where parameter ${\cal{F}}$ describes the fraction of time when the area $\rm d A$ has been under thunderclouds.

We present two different estimates for parameter ${\cal{F}}$ below. First 
estimation of parameter ${\cal{F}}$ is based 
on compilation of the annual thunderstorm duration from 450 air weather system in USA as described  in 
 \cite{Gurevich:2004km}.   The corresponding estimates suggest that on average the thunderstorms last about $1\%$ of  time in each given area  \cite{Gurevich:2004km}. If we adopt this estimate we arrive to conclusion that 
 \be
  \label{BL-flux1}
\frac{\rm d \Phi^{AQN}_{BL}}{\rm d A} 
 \approx  4\cdot 10^{-4}  \left(\frac{\rho_{\rm DM}}{0.3{\rm\,GeV\,cm^{-3}}}\right)
 \rm \frac{(BL~events)}{yr\cdot  km^2}, ~~~~  {\cal{F}}\approx 10^{-2}
\ee
     
 The second (independent)   estimation of parameter ${\cal{F}}$
 has been used  in our analysis \cite{Zhitnitsky:2020shd} of mysterious CR-like events, the  so-called Telescope Array bursts. This estimate  is based on the number of detected lightning events during  5 years (between May 2008 and April 2013) in the area. The corresponding number of lightning events   is 10073  \cite{Abbasi:2017rvx,Okuda_2019}.
   Assuming that a typical thunderstorm lasts  one hour and    produces $10^2$  lightnings     \cite{DWYER2014147} one can infer that the total time when  the  relevant area   area was under a  thunderstorm 
   is $10073\cdot  10^{-2} h\simeq  10^{2} h$ during 5 years of recording. This represents approximately fraction ${\cal{F}}\simeq 0.25\cdot 10^{-2}$. As a result we arrive to our second estimate for frequency of the BL events: 
\be
  \label{BL-flux2}
\frac{\rm d \Phi^{AQN}_{BL}}{\rm d A} 
 \approx   10^{-4}  \left(\frac{\rho_{\rm DM}}{0.3{\rm\,GeV\,cm^{-3}}}\right)
 \rm \frac{(BL~events)}{yr\cdot  km^2}, ~~~~  {\cal{F}}\approx 0.25\cdot 10^{-2}.
\ee
If efficiency of spallation which produces the  secondary $AQN_s$ from original AQN is less than 100\% the corresponding 
estimates (\ref{BL-flux1}) and (\ref{BL-flux2}) decrease correspondingly. If original AQN  produces (as  a result of spallation) more than one  
  $AQN_s$  than  the corresponding 
estimates (\ref{BL-flux1}) and (\ref{BL-flux2}) increase correspondingly.

With all these theoretical uncertainties mentioned above, we consider our estimates (\ref{BL-flux1}) and (\ref{BL-flux2}) are    perfectly consistent with lower bound (\ref{Phi_BL1}).
In fact, it is amazing that so different estimates which include dramatically different environments and physical systems 
(from thunderstorms and lightning events to dark matter density $\rho_{\rm DM}$ and galactic wind velocities explicitly entering all  the estimates) are so close to each other.

  Precisely the estimates (\ref{BL-flux1}) and (\ref{BL-flux2}) answer (within the AQN framework) the question formulated in the first paragraph of this section: why the BL events  are so rare? The proposed  answer is that  the rareness of  BL   events is a consequence of the rareness of the  AQN events with  very tiny DM  flux (\ref{Phi2}).  
  
  We conclude this section with the following comment. The consistency of our estimates (\ref{BL-flux1}) and (\ref{BL-flux2})   with lower bound (\ref{Phi_BL1}) along with multiple consistency checks summarized in Sect.\ref{sect:comments}    further strengthen the proposal (\ref{eq:proposal}) relating BL and DM physics.

\section{On Possible relation  between BL and  other Unidentified Aerial Phenomena (UAP)}\label{sect:UAP}

In this section we would like to present several arguments suggesting that some of the Unidentified Aerial Phenomena (UAP) might be also related to the DM physics. For a comprehensive review on UAP observations see recent review 
\cite{knuth2025newscienceunidentifiedaerospaceundersea}. 

 By obvious reasons this section is much more speculative in comparison with all our previous sections because 
     of scarcely real physical measurements of UAP events in comparison with sufficiently long list of  numerical estimates  for  BL events as reviewed  in Sect. \ref{items}. In fact, items (ix), (x) and (xi) from Sect. \ref{items} can be treated as   the  BL-features  recorded by modern physics instruments. This should be contrasted with UAP-related phenomena when very few observations could be considered as 
     well documented and  properly recorded events. In fact, scientific studies of the UAP phenomena  started very recently, see  e.g. papers   \cite{doi:10.1142/S2251171723400068,s25030783,Szydagis:2023lzo}.  Nevertheless, we opted to include  descriptions of several such observations into this work because  we believe that some of the UAP observations might be closely related to BL physics. 
 Therefore, some UAP events may also represent the manifestations of the DM physics  as a consequence of our   proposal (\ref{eq:proposal}).
     
      Indeed, there is one unique  but crucial common element  relating  BL phenomena and UAP: in both cases no  any material objects remain   in the active area where events took places. This is in spite of the fact that many secondary accompanying effects   had been observed and even measured in physics laboratories.   The secondary effects manifested in many different ways: in form of the visible light, X rays, shock acoustic waves (see Sect. \ref{sect:skyquakes}),  heat melting the glass or soil (see Sect.\ref{sect: ELMA}).  This is only possible when the source of the energy
   for both cases is represented by annihilation processes of the antimatter   with surrounding matter materials, which is precisely the key element of this proposal. 
     
     As a result of these similarities we propose that some of 
     the UAP events may be also considered as profound manifestations of the same  DM physics similar  to our proposal (\ref{eq:proposal}) relating BL and  DM physics.  Therefore, the main purpose of this section is to argue that many UAP events are in fact   very similar   to the  BL phenomena  discussed in Sect. \ref{sect:consistency}. The only difference  is that the UAP events   are characterized by dramatically different energy scales (larger baryon charge, and consequently much higher velocity  at the moment of  observation)  in comparison with BL events with low velocities  and     typical energies   (\ref{s-energy}).   
     
          \subsection{Pseudo-meteorites  events as the AQN-induced  events}\label{sect: ELMA}
          There are many reports in the literature describing the meteor-like events which  (after detail studies by professionals) turned out to be not the meteorites. The corresponding events are classified as ``Pseudo-meteorites  events", see   \cite{2020arXiv201200686O} with large number of  references and details. In this work we want to focus on a single event   which took place in US town of Elma (Washington State) on July 15, 2003 as described in \cite{2020arXiv201200686O}. Many other similar events discussed in  \cite{2020arXiv201200686O}   shall not be mentioned here, and  we think they   likely have the same nature, and are originated from the same physics. 
          
     The key elements of the Elma event are \cite{2020arXiv201200686O}: 1.after the event (fireball) the   glassy black rocks had been found; 2. the  rocks were very hot (one of the witnesses had burned a thumb and a finger by collecting the rocks); 3. analysis of the discovered stones by the University of Washington concluded  that the glassy rocks were not meteorites, see pictures  in  \cite{2020arXiv201200686O}; 4. security cameras did not record any suspicious images;  5. cloud of dust went up  in   the active area.  
     
     Our original comment is as follows. The Elma event is very similar to BL events discussed in Sect. \ref{sect:consistency} with the only difference is that the $AQN_s$ must be  much larger in size $B\gg 10^{15}$ in comparison with a typical BL event. In this case, as we discussed in item 3 in Sect. \ref{sect:stopping} the $AQN_s$
     cannot efficiently slowdown, and it   is likely to hit the earth's surface with a high speed. As a result,  the $AQN_s$  gets annihilated in deep underground regions (not in air, which is typical for BL events) by heating the surrounding rocks along its path. In many respects the heating of the surrounding material is similar to phenomena of the  BL passing through the glass window when the internal $AQN_s$ temperatures could reach enormous values due to much higher density of the soil in comparison with air, see our estimates  in Sect. \ref{sect:glass}.  It explains the hot glassy   rocks found in the area. Furthermore, for the $AQN_s$ 
     moving with a high speed the radiation  in visible frequency bands is suppressed (similar to ELFO event discussed in next Sect.\ref{sect:skyquakes}), in contrast with slow moving BL discussed in Sect. \ref{sect:radiation}. It explains the absence of any images by security camera. This  identification of the ``Pseudo-meteorites  events"   with $AQN_s$ events  is consistent with all observed puzzling observations listed in previous paragraph. Therefore, we speculate  that the ``Pseudo-meteorites  events"  along with BL events represent the  right hand side of the same identification  (\ref{eq:proposal}) and both types of  events represent   profound manifestations of the same DM physics. 
   
        \subsection{Sky-quakes as the AQN events}\label{sect:skyquakes}
      The so-called skyquakes  have been known for centuries, similar to BL events. They manifested in the form of sound and infrasound without leaving any   physical material  objects in the area in form of true meteorities.  Their nature   remains unknown in spite of the long history of observations, with records going back over 200 years. It has been speculated  in \cite{Budker:2020mqk} that
 skyquakes could be a manifestation of the dark matter AQN
 traveling in the atmosphere. 
 
\exclude{ Skyquakes \cite{skyquakes} are unexplained acoustic events that sound like a cannon shot or a sonic boom coming from the sky, see for example, a description in a TV interview by a meteorologist 
 \cite{skyquakes-meteorologist}. The main message of this short interview is  that skyquakes could not be due to any seismic events or meteor-type events which are routinely recorded around the globe. These events could not be identified with any human activities which are also recorded. One should also add that similar events have been recorded  for centuries in different countries with different  environmental features. These events cannot be explained by military aircraft as  records of skyquakes appeared long before supersonic flights. 
 }
Unfortunately, it is next to impossible to extract any useful quantitative information from the numerous but random and unsystematic records on sky-quakes collected for centuries (very similar to BL events). Luckily, one such event which occurred on July 31-st 2008 was properly recorded by the dedicated Elginfield Infrasound Array (ELFO) near London, Ontario, Canada, see   \cite{Budker:2020mqk}
for references and details.  The infrasound detection was accompanied by non-observation of any meteors by an all-sky camera network, ruling out a conventional meteor source. In addition, no any meteorites had been found in the active area,
similar to  Elma event mentioned above in Sect. \ref{sect: ELMA} and classified as ``Pseudo-meteorites  events". 
Anthropogenic sources such as operations at the nearby Bruce Nuclear Power Plant or the  Goderich salt mine were also eliminated; a  local airport radar reported no aircraft in the area at the time. 
%Furthermore, these events cannot be explained by military aircraft as  records of skyquakes appeared long before supersonic flights.  
In addition to infrasound, impulses were also observed seismically (few moments after the event) as ground-coupled acoustic waves around Southwestern Ontario and Northern Michigan. This event was treated as AQN-induced event as   discussed in detail in   \cite{Budker:2020mqk}
  where it was argued that    the energetics, infrasound-frequency properties, and other characteristics 
 of the event are consistent with observed ELFO event.
  
Our original comment here is as follows. As argued in  \cite{Budker:2020mqk} the skyquakes could be also AQN induced events. However, the scale of skyquakes are dramatically different from BL events and ``Pseudo-meteorites  events" mentioned above. In fact, it was estimated in  \cite{Budker:2020mqk} that the the baryon charge of ELFO event is around $B\approx 10^{27} $ which is two orders of magnitude above a typical AQN size\footnote{Such powerful events    are much more  rare ones than   typical AQN  events with rate given  by  (\ref{Phi1}) because the size distribution 
$f(B)\propto B^{-2}$} and many orders of magnitude above a typical $AQN_s$ scale. In different words,  skyquakes are the manifestations of the original AQN, not secondary $AQN_s$    with dramatically smaller sizes and masses (and consequently released energies). However, all  types of events discussed above (BL, Pseudo-meteorites,  and skyquakes) can be considered as different manifestations  of the same AQN-induced events when the source of the energy is the same, and it is hidden  in form of the antimatter nuggets  which had been produced   during the QCD epoch in early Universe.

%Such strong powerful and energetic events with 
%$B\approx 10^{27} $ are much more  rare than a typical BL rate given  by  (\ref{Phi2}) because the size distribution 
%$f(B)\propto B^{-2}$. 

\subsection{Unidentified Anomalous Vehicles (UAV) as manifestation of the AQN events}\label{sect:UAV}
This is the most speculative subsection of this work as no quantitative physical measurements exist of the so-called 
Unidentified Anomalous Vehicles (UAV) which have been observed globally, see   recent review 
\cite{knuth2025newscienceunidentifiedaerospaceundersea} and scientific publication  \cite{e21100939} with specific numerical estimates for the velocity and acceleration of UAV which seem to be  inconsistent with laws of physics if interpreted as aircraft. 

We interpret  the radar information  of the UAV image  from USS Nimitz nuclear aircraft carrier (14 November 2004),  see Fig. 5   in  \cite{e21100939}, as an image of a highly ionized cloud (cylindrical form) produced by passing AQN. In what follows we support this interpretation by estimation of the plasma frequency $\omega_p$ of this ionized cloud. In this case the pulse being sent from radar will be reflected  by this highly ionized cloud and recorded by USS Nimitz. The effect of reflection from ionized cloud is  similar to reflection of the AM radio bands from the Earth's  ionosphere. In contrast with ionosphere however where  $\omega_p\approx 2\cdot 10^{7} s^{-1}$ (MHz range),  the plasma frequency in  highly ionized cloud produced by passing AQN could easily assume GHz range.
As a result,   the AQN-induced cloud could be misinterpreted as   UAV image. In what follows we support our proposal  by providing  an order of magnitude  estimates for $\omega_p$ and the size of the  highly ionized cloud produced by passing AQN.  

Indeed, our order of magnitude estimates as given in Appendix \ref{sect:omega} suggest that the AQN propagating at high altitudes will be ionizing the surrounding area with very high efficiency. We estimated the plasma frequency for this ionized region to be on the level $\omega_p\approx 3\cdot 10^9 \rm s^{-1}$ or even much higher, according to (\ref{omega-final}).
This implies that the pulses with   GHz frequency bands from radar can be reflected by the ionized cloud\footnote{
One should emphasize that an ionized cloud  produced by passing AQN is the subject of atmospheric currents and can easily move and fluctuate.  The AQN itself entering the atmosphere with very high speed $\sim 10^{-3}c$ could be at a very different location at the  instant when the radar records the image of the ionized cloud.}, which can be misinterpreted as UAV. Furthermore, a typical size of the cloud is determined by the mean free path of the 20 keV photons emitted by very hot AQN. The corresponding mean free path is  estimated 
in  (\ref{lambda_z}) and it assumes the  numerical value    around $L_{\gamma}\approx \rm 15~ m$. This scale is   similar  to  the  size of   UAV image   mentioned in   ref. \cite{e21100939}. Finally, the acceleration and the velocity
of the AQN propagating in atmosphere have been estimated in  (\ref{eq:v(t)}) and (\ref{eq:acceleration}) and consistent
with  the values   presented in  ref. \cite{e21100939}. The estimation  (\ref{eq:acceleration}) shows that the acceleration could easily reach enormous magnitudes $\sim (10^2-10^4)g$ being inconsistent with modern technology   if interpreted as UAV according to ref. \cite{e21100939}. Of course, such acceleration  is perfectly consistent with physics laws if interpreted as   the AQN object which enters  the earth's atmosphere   with galactic velocity (around $10^{-3}c$)  and slows down as a result of collisions with   atoms and molecules in atmosphere as described in Appendix \ref{sect:stoppage}. 

We conclude this  section  by mentioning that  UAV along with BL, Pseudo-meteorites,  and skyquakes  as discussed above represent different manifestations  of the same AQN-induced events when the source of the energy is hidden  in the antimatter nuggets  which had been formed  during the QCD epoch in early Universe. All these events are very rare ones.  Their rareness is explained by very tiny DM flux formulated in terms of the well-constrained DM density $\rho_{\rm DM}$ as given by (\ref{Phi2}). 

It is obvious that we can not make any progress in understanding of the  UAP events mentioned above due to the lack of real data 
and proper physics measurements, which makes this section even more  speculative than   BL studies with  sufficiently long list of  numerical estimates  as reviewed  in Sect. \ref{items}.  Fortunately, the situation is about to change, and a number of projects and proposals, see e.g.     \cite{doi:10.1142/S2251171723400068,s25030783,Szydagis:2023lzo} have been put forward to fill this gap. The UAP    events in general and UAV events in particular can be and should be systematically studied. We   propose   in Sect. \ref{sect:UAP_tests}  several specific  tests on how  to substantiate or refute our proposal formulated above on identification   some of the UAP events  with AQN-induced phenomena.

\section {Concluding comments and Future Developments} \label{conclusion} 

 The presence of the {\it antimatter} nuggets\footnote{We remind the readers that the antimatter in this framework was suggested long ago  \cite{Zhitnitsky:2002qa, Zhitnitsky:2021iwg} as natural resolution of two fundamental  cosmological puzzles: 1. similarity  between visible and DM components, $\Omega_{\rm DM}\sim \Omega_{\rm visible}$; 2. observed baryon asymmetry of our Universe. These puzzles are automatically resolved in the AQN framework  irrespective to the parameters of the model. A mechanism of formation of these antimatter nuggets is reviewed in Sect. \ref{AQN}. }  
 in the system implies, as reviewed in Sect.\ref{AQN}, that there will be  annihilation events   leading to enormous energy injection into surrounding region. As a result of these annihilation events one should anticipate a  large number of observable effects on different scales: from Early Universe to the galactic scales to the Sun and the  terrestrial   events. 

 In the present  work we focused on manifestations of these annihilation events on possible  resolution of the mysterious BL physics. In next Section \ref{sect:tests} we suggest specific  tests which could  support or refute the proposed resolution of these BL-related mysteries. In Section  \ref{sect:UAP_tests} we suggest different types of  experiments 
with the  instruments  which could support or refute our suggestion that some  UAP events (such as Pseudo-meteorites,  skyquakes, UAV) could be also related to the AQN-induced phenomena
 as argued in Sect.\ref{sect:UAP}.  Finally,  in Sect.  \ref{sect:paradigm} we overview other manifestations of the same AQN framework  at dramatically different scales: from   the early Universe to galactic scales,  to the solar scale  to the Earth scale,  where similar mysterious puzzles are known for decades (even centuries), and  could be also related to  the same  antimatter nuggets (representing the DM objects) within the same AQN framework.
 We emphasize  that the    parameters of the model were 
 fixed long ago irrespective to the BL or UAP  physics which represents  the topic of the present work.
   
          \subsection{BL as manifestation of the $AQN_s$ events. Possible future tests of  the proposal (\ref{eq:proposal}).}\label{sect:tests}
          We already formulated the basic results demonstrating the consistency of the proposal (\ref{eq:proposal})
          with observed BL features   in  Sect.        \ref{sect:comments}. We do not need to repeat these results again.
          Instead, we focus here  on possible future tests of the proposal (\ref{eq:proposal}). Before we formulate 
          the basic idea for  the test we would like to make few comments on observed  anomalous and mysterious cosmic rays (CR)-like events 
          which are very hard to explain within conventional framework and   modelling. These events are strongly associated 
          with thunderstorm and lightning events in the area, similar to BL events discussed in this work.
          
           In particular,  the Telescope Array  (TA) collaboration  \cite{Abbasi:2017rvx,Okuda_2019} reported the observations of   mysterious   bursts    when 
   at least three air showers  were recorded within 1 ms which cannot occur with conventional high energy CR (it should be days or event months for two or more consecutive  energetic CR events hit the same area).  These TA mysterious   bursts are associated with lightning events in the area. We proposed  in   \cite{Zhitnitsky:2020shd} that these puzzling  events  could be related to the AQN-induced events.
   In fact, in ref. \cite{Zhitnitsky:2020shd} we used the same formula (\ref{Phi2}) for estimation of the  rate of 
    TA mysterious   bursts  which was  used for estimation of the BL frequency of appearance in (\ref{BL-flux2}).  We explained the rareness of these TA mysterious   bursts   precisely   in the same way as  the rareness of BL events.

   Similar ``Exotic Events" recorded by  the  AUGER   collaboration  \cite{PierreAuger:2021int,2019EPJWC.19703003C,Colalillo:2017uC} are also associated with thunderstorm activity in the area. These events also cannot be explained by canonical CR modelling. At the same time, 
   these ``Exotic Events" recorded by  the  AUGER   collaboration   can be   explained within AQN framework as argued in \cite{Zhitnitsky:2022swb}. Furthermore, the rareness of these ``Exotic Events" is also explained in the AQN framework by 
   same formula (\ref{Phi2}) and it is consistent with counting of the ``Exotic Events" by the  AUGER   collaboration. 
   
   We mentioned these two studies on unusual CR like events to  propose to   test the hypothesis (\ref{eq:proposal}) as follows.
   If one can install all sky cameras, similar to the ones used in  analysis \cite{BL-spectrum-1}, to monitor entire  sky  in the same area where TA or AUGER detectors  (or any other CR labs) are  located   one can record the BL events with frequency of appearance   similar to   mysterious   bursts recorded by TA collaboration or ``Exotic Events" recorded by AUGER   collaboration\footnote{Similar all sky camera had been used by monitoring 
  entire  sky in connection with studies of the meteoroids, and searches for correlations with infrasound signals, see 
 Section  \ref{sect:skyquakes}.}. Indeed, in all cases the basic formula describing the rate for all these events is (\ref{BL-flux2}).  This formula is shown to be consistent with TA mysterious   bursts (10 events recorded during 5 years),  AUGER ``Exotic Events" (23 events recorded during 13 years). The same formula is also consistent with lower bound for BL events  (\ref{Phi_BL1}) as discussed in Sect. \ref{sec:event_rate}.
 Therefore, we predict (within AQN framework) that the rate  for BL events in the area will be similar to observed TA mysterious   bursts events and AUGER ``Exotic Events" because all these events are different manifestations of the same DM physics.    
 
 Furthermore, we also predict, within the same AQN framework, that TA mysterious   bursts will be correlated in time with observations of the BL in the same area under thunderstorm. The same comment obviously applies to AUGER ``Exotic Events". The recording of such correlation would be very strong and unambiguous argument supporting entire idea about the nature of BL as profound manifestation of the DM physics, as advocated in this work.
 
   \subsection{UAP as manifestation of the AQN events. Possible future tests.}\label{sect:UAP_tests}
   
   In this subsection we would like to make few comments on possible tests of our interpretation of some UAP events  as discussed in Section \ref{sect:UAP} as the direct manifestation of the AQN events, similar to our interpretation of the BL
   events as the $AQN_s$ events.  However, in contrast with proposed tests in previous subsection \ref{sect:tests}
   the required instruments to detect the signals from UAP should be  very different as they must be  sensitive to very different frequency bands. 
   
   Therefore, based on the discussions in Sections \ref{sect:skyquakes} and \ref{sect:UAV} we suggest to test our proposal
   relating BL, Pseudo-meteorites, skyquakes and UAV phenomena with AQN-induced effects (representing the DM physics in this framework) as follows. First of all, one can search for  infrasound acoustic signals, similar to our suggestion formulated in    \cite{Budker:2020mqk} where we proposed to use Distributed Acoustic Sensing (DAS), which is becoming a conventional tool for seismic and other applications. The basic idea of these activities  can be explained  as follows.  It has been known for quite sometime that distributed optical fiber sensors are  capable of measuring the signals  at thousands of points simultaneously using  an unmodified optical fiber as the sensing element. The recent  development is that the DAS is capable of measuring strain changes at all points along the optical fiber at {\it acoustic frequencies}, which is crucial for our studies of the acoustic waves emitted due to the AQN passage.
   
   Another element which  could play a crucial role in identification of the UAV is the radar. We suggest to use two different frequencies simultaneously, one is with larger than plasma frequency, $\omega> \omega_p$  and another one is with smaller than plasma frequency $\omega< \omega_p$. The radar with $\omega< \omega_p$ will detect the image of the ionized cloud produced by  the passing AQN, while another radar with   $\omega> \omega_p$ will be blind to the same area. 
   The recording of the  correlation between two radars with different frequencies and DAS  should  be considered as a very strong and unambiguous argument supporting entire idea about the nature of UAP  as profound manifestation of the DM physics in form of the antimatter nuggets powering all these mysterious events. The event rate is determined by the DM flux (\ref{Phi2}) which is consistent with the BL lower bound as discussed in Section \ref{sec:event_rate}, and consistent with observed mysterious and anomalous events (which we believe are consequences of  the AQN events) recorded by AUGER and Telescope Array collaborations as reviewed in Section \ref{sect:tests}.

     \subsection{Other (indirect) evidences for  DM in form of the AQN}\label{sect:paradigm}

  There are many hints  suggesting that the  annihilation events and consequent energy injection   into space (which is inevitable feature of this framework) may  indeed  took place in early Universe, during the galaxy formation as well as   in present epoch. We would like to mention 
  (for completeness of the presentation) a number of mysterious puzzles at different scales which could be also 
  related to the additional energy injection induced by the AQN annihilations events with surrounding  material.  
  
 We start this list of (yet) unresolved puzzles  from early Universe epoch, more specifically with BBN when the so-called the  ``Primordial Lithium Puzzle" has been with us for decades.
It has been argued in  \cite{Flambaum:2018ohm} that  the  AQNs during the BBN epoch do not affect BBN production for H and He, but   might be responsible for a resolution of   the  ``Primordial Lithium Puzzle" due to  the Li    large electric charge $Z=3$    strongly interacting with negatively charged AQN.
%\footnote{The AQN will be ionized as a result of high temperature during this epoch when many weakly bound positrons from the AQN's electrosphere leave the system such that AQN becomes strongly negatively charged object during BBN epoch.}  
  
  Another well known puzzle is related to the galaxy formation epoch. The corresponding puzzles commonly are formulated as  ``Core-Cusp Problem", ``Missing Satellite Problem", 
  ``Too-Big-to-Fail Problem", to name just a few, see recent reviews 
 \cite{Tulin:2017ara,Salucci:2020eqo} for the details\footnote{\label{puzzles}There are many more similar problems and very puzzling observations. We refer to the review papers  \cite{Tulin:2017ara,Salucci:2020eqo} on this matter. There are also  different, but related  observations which  apparently  inconsistent with conventional picture of the structure formation \cite{Salucci:2020eqo}.}. It has been argued in \cite{Zhitnitsky:2023znn} that   the aforementioned discrepancies  (and many other related problems referred to  in footnote \ref{puzzles})  may be alleviated if dark matter is represented in form of the  composite,    nuclear density objects within AQN framework. 
  
  We now move from the early times in evolution of the Universe to the present day observations.  In this case there is  a  set of puzzles which is related to  the  diffuse UV emission in our galaxy. 
In has been claimed  in  \cite{Henry_2014,Akshaya_2018,2019MNRAS.489.1120A} that there are many observations which 
   are very hard to understand if interpreted in terms of the conventional astrophysical  phenomena.
 The  analysis \cite{Henry_2014,Akshaya_2018,2019MNRAS.489.1120A}  very convincingly disproves   the conventional picture  that  the dominant source of the diffused  UV background  is the dust-scattered radiation of the UV emitting stars.   
 The arguments are based on a number of very puzzling observations which are not compatible with standard picture. First, the diffuse radiation is very uniform in both hemispheres, in  contrast to the strong non-uniformity in distribution of the UV emitting stars.  Secondly, the diffuse radiation is almost entirely independent of Galactic longitude. This feature must be contrasted with localization of the brightest UV emitting stars which are overwhelmingly confined to the longitude range $180^0-360^0$. 
 These and several similar observations   strongly suggest that the diffuse background radiation can hardly  originate in dust-scattered starlight.  The authors of \cite{Henry_2014} conclude that the  source of the diffuse FUV emission is unknown --that is the mystery that is referred to in the title of the paper \cite{Henry_2014}.

It has been proposed  in  \cite{Zhitnitsky:2021wjb}  that  this excess in UV radiation  could be a  result of the 
    dark matter annihilation events  within  the AQN dark matter 
    model.   The proposal  \cite{Zhitnitsky:2021wjb}  is supported by demonstrating that   intensity and the spectral features of the   AQN induced emissions    are consistent with  the corresponding  characteristics of the observed excess  \cite{Henry_2014,Akshaya_2018,2019MNRAS.489.1120A}  of the UV radiation.

  We move from galactic to solar scale.   The AQNs might be also responsible for renowned    long standing problem\footnote{This persisting puzzle is characterized by the following observed anomalous behaviour of the Sun:
the quiet Sun (magnetic field $B\sim 1$  Gauss) emits an extreme ultra violet (EUV) radiation  with a photon energy of order {\rm $10^2~eV $ } which cannot be explained in terms of  any conventional astrophysical phenomena. %At the transition region, the (quiet Sun) temperature continues to rise very steeply  until it reaches  a few $10^{6}$~K, i.e., being a few 100 times hotter above the underlying  photosphere, 
This happens within an atmospheric layer thickness of only 100 km or even much less.  The variation of EUV with solar cycles is very modest and of order of (20-30)\%  during the solar cycles when magnetic activity varies by factor $10^2$ or more. So, it is hard to imagine how the  magnetic reconnection, which is known to be responsible for large flares, could play any role  when $B\sim 1$  Gauss. There are many other puzzling features discussed in \cite{Zhitnitsky:2017rop,Raza:2018gpb}.} of  the ``Solar Corona Mystery"
  when the   so-called ``nanoflares" conjectured by Parker long ago \cite{Parker} are   identified with the  annihilation events in the AQN framework\cite{Zhitnitsky:2017rop,Raza:2018gpb}.

    We move from solar scale to terrestrial unusual events. We already mentioned mysterious CR like  events in Sect. \ref{sect:tests}. There are many other terrestrial unusual events   when conventional picture cannot explain the observed phenomena. In particular,    the   \textsc{ANITA}  observed      two anomalous events    with non-inverted polarity   \cite{Gorham:2016zah,Gorham:2018ydl}.  Such events correspond to very large inclination angle  when ``something" crossing the earth before exiting from opposite side of earth  at the moment of  recording by \textsc{ANITA}. Such events  are very hard to explain within conventional physics, but could be  explained within AQN framework \cite{Liang:2021rnv}.  
    % the  Multi-Modal Clustering  Events observed by HORIZON 10T \cite{2017EPJWC.14514001B,Beznosko:2019cI}. It is also  hard to understand  in terms of the conventional CR events, but    could be  interpreted in terms of  the  AQN annihilation events in atmosphere   as argued in  \cite{Zhitnitsky:2021qhj}. 

Finally, it has been recently argued in \cite{Zioutas_2020,Argiriou:2025xsq} that numerous enigmatic observations remain challenging to explain within the framework of conventional physics.  In particular, these anomalies include unexpected correlations between temperature variations in the stratosphere and  the total electron content of the Earth's atmosphere
(along with many other mysterious correlations), such that entire globe can be thought as a one large detector.  Decades of collected data provide statistically significant evidence for these observed correlations.  However, no any conventional explanations for such mysterious correlations had been offered.  It    has been argued in \cite{Zhitnitsky:2024jnk}  that the recorded correlations can be 
generated by the AQN-induced processes.

          \vspace{0.5cm}
          
   We conclude this work with the following final comment. 
   We advocate an idea that   the BL known for centuries might be a profound manifestation of the DM physics  when DM  is  made  of (anti)quarks and gluons 
   of the Standard Model as reviewed in Sect.\ref{AQN}. We also speculate that other mysterious phenomena known as UAP (it includes pseudo-meteorites,  skyquakes and  UAV) might   also represent different manifestations 
   of the same AQN dark matter physics. The AQN dark matter model   is  consistent with all presently available cosmological, astrophysical, satellite and ground-based observations.   In fact, it may even shed some light  on the  long standing puzzles and mysteries as briefly reviewed above in Section \ref{sect:paradigm}.    If validated through the proposed tests and experiments, this framework could revolutionize our understanding of DM and its role in the cosmos. The BL and UAP events as   profound direct manifestations of the DM on earth (within the AQN framework)  could play a   key role  in   the  {\it direct} study of the DM   in scientific labs, in contrast  with {\it indirect} manifestations of the AQN-induced phenomena reviewed in Sect. \ref{sect:paradigm}.

     \section*{Acknowledgements}
      This research was supported in part by the Natural Sciences and Engineering
Research Council of Canada.

\appendix
  \section{AQN emission spectrum}\label{sect:spectrum} 
 The goal of this Appendix is to overview the spectral characteristics of the AQNs as a result of annihilation events 
 when the nugget enters the Earth atmosphere. The corresponding computations have been carried out in \cite{Forbes:2008uf}
 in application to the galactic environment with a typical density of surrounding visible baryons of order $n_{\rm galaxy}\sim 1 ~{\rm cm^{-3}}$ in the galaxy. We review  these computations with few additional elements which must be implemented for Earth's atmosphere when 
 typical density of surrounding baryons is much higher $n_{\rm air}\sim 10^{21} ~{\rm cm^{-3}}$.
 
The spectrum of nuggets at low 
temperatures was analyzed in \cite{Forbes:2008uf} and was found to be,
\be 
  \label{eq:spectrum}
  \frac{d{F}}{d{\omega}}(\omega) = 
  \frac{d{E}}{d{t}\;d{A}\;d{\omega}}
  \simeq
  \frac{1}{2}\int^{\infty}_{0}\!\!\!\!\!d{z}\;
  \frac{d{Q}}{d{\omega}}(\omega, z)
  \sim \\
  \sim
  \frac{4}{45}
  \frac{T^3\alpha^{5/2}}{\pi}\sqrt[4]{\frac{T}{m}}
  \left(1+\frac{\omega}{T}\right)e^{-\omega/T}h\left(\frac{\omega}{T}\right),
\ee
where $\alpha=137^{-1}$ is the fine structure constant and we use units $\hbar=c=k_B=1$. The function $Q(\omega, z)\sim n^2(z, T)$ describes the emissivity   per unit volume from the electrosphere   characterized by the density  $n(z, T)$, where $z$ measures the distance from the quark core of the nugget. The $\frac{d{F}}{d{\omega}}(\omega)$ describes the intensity of emission from unit area $A$ at frequency $\omega$ at temperature $T$.   In Eq.\,(\ref{eq:spectrum}) a complicated function $h(x)$ can be well approximated as 
\begin{equation}
  h(x) = \begin{cases}
    17-12\ln(x/2) & x<1,\\
    17+12\ln(2) & x\geq1.
  \end{cases}
\end{equation}
Integrating over $\omega$ contributes a factor of
$T\int d{x}\;(1+x)\exp(-x)h(x)\approx 60\,T$, giving the total surface
emissivity:
\begin{equation}
  \label{eq:P_total}
  F_{\text{tot}} = 
  \frac{d{E}}{d{t}\;d{A}} = 
  \int^{\infty}_0\!\!\!\!\!d{\omega}\;
  \frac{d{F}}{d{\omega}}(\omega) 
  \sim
  \frac{16}{3}
  \frac{T^4\alpha^{5/2}}{\pi}\sqrt[4]{\frac{T}{m}}.\\
\end{equation}

 A typical internal temperature of  the nuggets can be estimated from the condition 
 the radiative output of equation (\ref{eq:P_total}) must balanced the flux of energy onto the 
nugget  due to the annihilation events. In this case one arrives to  equation (\ref{eq:rad_balance}) from the main body of the text. 
 
The factor $\kappa$ is introduced to account for  complicated physics as mentioned in the main body of the text. 
 In a neutral environment when no 
long range interactions exist the value of $\kappa$ cannot exceed $\kappa \sim 1$ which would 
correspond to the total annihilation of all impacting matter into to thermal photons. The high probability 
of reflection at the sharp quark matter surface lowers the value of $\kappa$. The propagation of an ionized (negatively charged) nugget in a  highly ionized plasma    will 
 increase the effective cross section, and therefore value of  $\kappa$ as discussed in \cite{Raza:2018gpb} in application to the solar corona heating problem. 
 
 For the neutral environment (such as Earth's atmosphere) and relatively low temperature when the most positrons from electrosphere remain in the system  the parameter $\kappa$ should assume values close to unity, i.e. $0.1\leq\kappa\leq 1$.  In this case, from (\ref{eq:rad_balance})
 assuming that $0.1\leq\kappa\leq 1$ one arrives to the estimate (\ref{T}) from the main  body of the text.
 
  There are few additional elements which should be taken into account for Earth's atmosphere in comparison with  original computations \cite{Forbes:2008uf,Forbes:2009wg} applied to very 
dilute galactic environment with much lower temperatures $T\simeq 1 $ eV. 
However, these effects in general do not modify the basic scale used in the main body of the  text (\ref{T}).
 
 For our analysis in Sect.  \ref{sect:consistency} we need  to estimate the internal temperature of the $AQN_s$ after its complete  stop when the rate annihilation is determined by  typical velocities  of molecules in air (\ref{eq:v_air}) rather by $v_{\rm AQN}$ itself. Due to dramatic decrease of the annihilation rate the equilibration internal temperature also decreases in comparison with (\ref{T}), and  can be estimated as follows:   
 \be
 \label{T-spallation}
   T\sim 20 ~{\rm keV} \cdot \left(\frac{n_{\rm air}}{10^{21} ~{\rm cm^{-3}}}\right)^{\frac{4}{17}}\cdot \left(\frac{v_{\rm air}}{v_{\rm AQN}}\right)^{\frac{4}{17}}\cdot 4^{\frac{4}{17}}\approx 6~\rm  keV,  
 \ee
where   factor $4^{\frac{4}{17}}$   accounts for the difference between geometrical cross section $\pi R^2$ for fast moving AQN and $4\pi R^2$ for $AQN_s$ at rest when air molecules from all angles can hit the $AQN_s$. Precisely this internal temperature is our benchmark value to be used in the main body of the text. 

 For our analysis in Sect.  \ref{sect:consistency} we also   need the estimate for the effective size $R_{\rm eff}$ of the AQN$_s$ after spallation. We define  the effective radius  $R_{\rm eff}$ as the scale where positrons from electrosphere remain to be strongly bound to the quark's core  at the internal temperature $T\approx 6$ keV.  The corresponding scale can be estimated from the   condition that the binding energy of the positrons is approximately equal to internal temperature $T$ of the system, similar to our previous estimates for the galactic environment  \cite{Forbes:2008uf,Forbes:2009wg}, i.e.
 \be
 \label{R_eff}
\rm  \frac{\alpha Q_{\rm AQN_s}}{R_{\rm eff}}\approx T\approx 6 ~ keV~~~~ \rightarrow ~~~~R_{\rm eff}\approx 10^{-6} cm, ~~~~~~~~ [ to~ be~ identified~ with~ internal~ size ~of~an~ AQN_s]
\ee
 where we use estimate (\ref{s-charge}) for $Q_{\rm AQN_s}$. As expected $R_{\rm eff}$ is dramatically larger than $R\approx 10^{-8} \rm cm$ from (\ref{s-energy}) corresponding to the size  of the quark matter core itself. One should emphasize that the physical meaning of the scale $R_{\rm eff}$ is very different from the scales discussed in Sects. \ref{sect:radiation} and \ref{sect:glass}. The $R_{\rm eff}$ literally describes the internal structure of the $AQN_s$ 
 as the positrons from electrosphere are strongly bound to the quark core. It must be contrasted with the scales from Sects.   \ref{sect:radiation} and \ref{sect:glass} which describe the mean free paths of the photons emitted from 
 electrosphere. These structures (including visible size of BL) emerge as a result of the interaction with surrounding material.  In other words, these scales are determined by the environment  where $AQN_s$  propagates, rather than by internal structure of $AQN_s$ itself.

\section{  The BL's  stoppage time   from the instant  of formation   to the moment of    observation }\label{sect:stoppage}
The main goal of this  Appendix is to provide a quantitative description of the kinematical motion of the  $AQN_s$ from the instant  of formation in  thunderclouds at high altitudes to the moment when it starts to behave as  a typical  BL object emitting the  visible frequency light  close to the  Earth's surface. We also 
justify our estimate as given in Sect. \ref{sect:stopping} on typical time scale when BL becomes  observable  in visible frequency bands.

The process of slowing  down of a heavy AQN  entering the atmosphere has been discussed previously in the context of  the solar corona in  \cite{Raza:2018gpb}. However, it can be directly applied to the $AQN_s$ propagating in the Earth's atmosphere which is the topic of the present work.  
Therefore, we can generalize    formula from \cite{Raza:2018gpb} to describe the  anti-matter  $AQN_s$    moving with velocity $v_{\rm AQN_s}(t)$ in   the atmosphere with matter density $\rho_{\rm air}\approx m_p n_{\rm air}$:
 \be
\label{eq:rate2}
  M_{\rm AQN_s}   \frac{d v_{\rm AQN_s}(t)}{dt}\approx   -\frac{\pi R_{\rm eff}^2}{2} \cdot  (m_p n_{\rm air}) \cdot v^2_{\rm AQN_s}(t),  ~~~~ M_{\rm AQN_s} \approx m_p B_s.
\ee 
 In what follows we assume that $B_s$ varies with time much slower than $v_{\rm AQN_s}(t)$, in which case $B_s$ can be treated as  constant initial baryon  charge entering estimate (\ref{eq:stop}) for $L$. In this case the equation (\ref{eq:rate2}) can be simplified as follows:
 \be
\label{eq:rate3}
     \frac{d v_{\rm AQN_s}(t)}{dt}\approx   -\frac{1}{2L} \cdot  v^2_{\rm AQN_s}(t).    
     \ee 
 With all these assumptions and approximations the solution for velocity   $v_{\rm AQN_s}(t)$ as a function of time $t$
 assumes the form
  \be
\label{eq:v(t)}
    \left( \frac{1}{v_{\rm AQN_s}(t)} -\frac{1}{v_{\rm AQN_s}(t_0)}\right) \approx   \frac{   (t-t_0)}{2L}, ~~~~{\rm where}~~~ v_{\rm AQN_s}(t_0)\approx v_{\rm AQN}(t_0)\approx 10^{-3}c
     \ee
 Our estimate (\ref{eq:v(t)}) suggests that a typical stoppage time scale $\tau_s$ when velocity $v_{\rm AQN_s}(\tau_s)$ 
 assumes only $\sim 1\%$ of its initial value after spallation can be estimated as follows
   \be
\label{eq:tau_s}
    \tau_s\approx \frac{2L}{ v_{\rm AQN_s}(\tau_s)}\approx \frac{2L}{[0.01\cdot v_{\rm AQN_s}(t_0)]}\approx \frac{6~ \rm km}{10^{-5}c}\approx 2 \rm~ s,
        \ee
 which is precisely the value being used in the main body of the text in Sect. \ref{sect:stopping}. 
 The main argument to define stoppage time scale $\tau_s$  as the scale  when velocity $v_{\rm AQN_s}(\tau_s)$ 
approximately  assumes   $\sim 1 \%$ fraction  from its  original value $300~ \rm km/s$  (when DM object enters the atmosphere)  is that    the internal temperature of the nugget at this velocity is close  to $T\approx 6~ \rm keV$. Precisely at this typical internal temperatures the     BL starts to radiate efficiently in the  visible  frequency bands  as discussed in Sect. \ref{sect:radiation}. 

Indeed, internal temperature $T$ at the  velocity $[0.01\cdot v_{\rm AQN_s}]$ assumes the form
 \be
 \label{T-spallation1}
   T\sim 20 ~{\rm keV} \cdot \left(\frac{n_{\rm air}}{10^{21} ~{\rm cm^{-3}}}\right)^{\frac{4}{17}}\cdot 
   \left(\frac{0.01\cdot v_{\rm AQN_s}(t_0)}{v_{\rm AQN}(t_0)}\right)^{\frac{4}{17}} \approx 6.7~\rm  keV,
 \ee
 and becomes very similar to  the temperature  (\ref{T-spallation}) we used in our analysis in  Sect. \ref{sect:radiation}.
 
 The equation (\ref{eq:v(t)}) allows us to estimate a typical acceleration of the BL after spallation time at $t_0$ as follows:
 \be
 \label{eq:acceleration} 
a_{v_{\rm AQN_s}}\equiv \frac{{v_{\rm AQN_s}(t)}}{dt}\approx -\frac{2L}{(t-t_0)^2}\approx 1.5\cdot 10^3\rm \frac{m}{s^2}\cdot \left(\frac{\tau_s}{(t-t_0)}\right)^2\gg 10^2 g.
 \ee
 The estimate (\ref{eq:acceleration}) shows that the acceleration of the $AQN_s$ could be enormous and could easily  exceed $10^2 $g as a result of very strong friction as it propagates in atmosphere. In fact it could easily exceed $10^4$g at $(t-t_0)\approx 0.1 \tau_s\approx 0.2 s$ when the velocity of the $AQN_s$ already dropped by 90\% from  its original value to reach $30~ \rm km/s$. Such enormous acceleration 
 had been mentioned previously in the literature \cite{e21100939} in relation with Unidentified Aerial Phenomena (UAP).
 We discuss a possible relation between BL and UAP phenomena in the main body of the text in Sect. \ref{sect:UAP}.

\section{On   ionization  of  air along the AQN path}\label{sect:omega}
The main goal of this appendix is to estimate the plasma frequency $\omega_p$ which is the basic element of the discussions of Sect. \ref{sect:UAV}. 
 The plasma frequency $\omega_{p}$   characterizes the propagation of photons in the ionized plasma.  The  $\omega_{p}$ can be thought as an effective mass for photon: only photons (from radar) with the energy larger than this mass can propagate in ionized plasma, while  photons with $\omega<\omega_{p}$  will be reflected by ionized plasma back to the radar producing the corresponding image of the ionized cloud.  The relevant formula for $\omega_p$ reads: 
\begin{equation}
\label{eq:omega}
    \omega_p^2=\frac{4\pi\alpha n_{\rm ion}}{m}.
\end{equation}
 To estimate $\omega_p$ we have to estimate the density of the ions $n_{\rm ion}$ which will be produced along the AQN path. 
 The corresponding estimates had been performed in  \cite{Budker:2020mqk} and we use formula (9) from that work
  \be
\label{E_l}
 E_l  
 \simeq 10^4\cdot   \left( \frac{B}{10^{25}}\right)^{2/3} \left(\frac{n_{\rm air}}{10^{21} ~{\rm cm^{-3}}}\right)\frac{\rm J}{\rm m}, 
   \ee
   where $E_l$ has the physical meaning of the energy per unit length injected along  the AQN's path.  The energy injection in \cite{Budker:2020mqk} was estimated at the sea level. For energy injection at high altitudes $z$ one should insert factor $\exp{\left( -{z}/{h}\right)}$ accounting for density decrease with  altitude, see (\ref{lambda_z}) below.  After that one can estimate the number of emitted photons per unit length  along  the AQN's path by dividing  $E_l$ to the internal  temperature of the nugget, i.e. $E_l/T$. 
   
   The next step is to estimate the photon's mean free path $ L_{\gamma}$ at high altitude $z\approx 8.5 \rm ~km $ where UAV had been observed. It has been also  computed in   \cite{Budker:2020mqk} for the sea level altitude and we use formula (12) from that work with adjustment   for the altitude dependence:
      \be
   \label{lambda_z}
   L_{\gamma}(z)=\frac{\lambda}{\rho_{\rm air}(z)} \simeq 5{\rm m}\cdot \exp{\left(\frac{z}{h}\right)} \approx 15 {\rm m}, ~~ {\rm where} ~~~h\approx 8 {\rm ~km}.
   \ee
The next step is to estimate the density of the ions $n_{\rm ion}
$ along the AQN propagation's path  as follows:
\be
\label{ion}
n_{\rm ion}\approx \frac{E_l(z)}{[\pi L_{\gamma}^2(z)\cdot T]}\approx 2\cdot 10^9 \rm \frac{ions}{cm^3}, 
\ee
where we assume that every $T\approx 20\rm ~ keV$ photon (which is our benchmark value) produces a single ion (through the photo-effect as discussed in Sect. \ref{sect:radiation}). We consider this estimate as a very conservative assumption as $  20\rm ~ keV$  injected energy is capable to produce multiple  ions through the chain of processes, such that the $n_{\rm ion}$
could  be in fact much higher than estimate (\ref{ion}) suggests. 

The final step in our estimate is to substitute (\ref{ion}) into expression (\ref{eq:omega}) for the plasma frequency
to arrive 
\begin{equation}
\label{omega-final}
    \omega_p=\sqrt{\frac{4\pi\alpha n_{\rm ion}}{m}}\approx 3\cdot 10^9 \rm s^{-1},
\end{equation}
which corresponds to the GHz range. This estimate is used in the main body of the text in Sect. \ref{sect:UAV}.

  \bibliography{BL-lightning}

\end{document}